\documentclass[fleqn,usenatbib]{mnras}

\usepackage{newtxtext}
\usepackage[varvw]{newtxmath}

\usepackage[T1]{fontenc}

\DeclareRobustCommand{\VAN}[3]{#2}
\let\VANthebibliography\thebibliography
\def\thebibliography{\DeclareRobustCommand{\VAN}[3]{##3}\VANthebibliography}

\usepackage{diffcoeff}
\usepackage{cleveref}
\usepackage{float}
\usepackage{placeins}

\usepackage{scalerel}
\usepackage{tikz}
\usetikzlibrary{svg.path}
\definecolor{orcidlogocol}{HTML}{A6CE39}
\tikzset{
  orcidlogo/.pic={
    \fill[orcidlogocol] svg{M256,128c0,70.7-57.3,128-128,128C57.3,256,0,198.7,0,128C0,57.3,57.3,0,128,0C198.7,0,256,57.3,256,128z};
    \fill[white] svg{M86.3,186.2H70.9V79.1h15.4v48.4V186.2z}
                 svg{M108.9,79.1h41.6c39.6,0,57,28.3,57,53.6c0,27.5-21.5,53.6-56.8,53.6h-41.8V79.1z M124.3,172.4h24.5c34.9,0,42.9-26.5,42.9-39.7c0-21.5-13.7-39.7-43.7-39.7h-23.7V172.4z}
                 svg{M88.7,56.8c0,5.5-4.5,10.1-10.1,10.1c-5.6,0-10.1-4.6-10.1-10.1c0-5.6,4.5-10.1,10.1-10.1C84.2,46.7,88.7,51.3,88.7,56.8z};
  }
}

\newcommand\orcidicon[1]{\href{https://orcid.org/#1}{\mbox{\scalerel*{
\begin{tikzpicture}[yscale=-1,transform shape]
\pic{orcidlogo};
\end{tikzpicture}
}{|}}}}

\usepackage{hyperref}

\usepackage{graphicx}	
\usepackage{amsmath}	

\title[Enhanced Phase Mixing of Torsional Alfv\'en Waves]{Enhanced Phase Mixing of Torsional Alfv\'en Waves in Stratified and Divergent Solar Coronal Structures, Paper I: Linear Solutions}

\author[C. Boocock and D. Tsiklauri]{
C. Boocock$^1$\thanks{E-mail: c.boocock@qmul.ac.uk} \orcidicon{0000-0002-8080-0555} and D. Tsiklauri$^2$ \orcidicon{0000-0001-9180-4773}
\\
$^1$Queen Mary University of London, Mile End Road, London E1 4NS, UK, \\
$^2$School of Science and Technology,
University of Georgia,
77a Kostava Street,
0171 Tbilisi,
Georgia
}

\date{Accepted 2021 November 26. Received 2021 November 25; in original form 2021 October 20}

\pubyear{2021}

\begin{document}
\label{firstpage}
\pagerange{\pageref{firstpage}--\pageref{lastpage}}
\maketitle

\begin{abstract}
We derive a corrected analytical solution for the propagation and enhanced phase mixing of torsional Alfv\'en waves, in a potential magnetic field with exponentially divergent field lines, embedded in a stratified solar corona. Further we develop a code named \textit{TAWAS} which calculates the analytic solution describing torsional Alfv\'en waves using IDL software language. We then use \textit{TAWAS} to demonstrate that both our correction to the analytic solution and the inclusion of wave reflection have a significant impact on Alfv\'en wave damping. We continue to utilise \textit{TAWAS} by performing a parameter study in order to identify the conditions under which enhanced phase mixing is strongest. We find that phase mixing is the strongest for high frequency Alfv\'en waves in magnetic fields with highly divergent field lines and without density stratification. We then present a finite difference solver, \textit{Wigglewave}, which solves the linearised evolution equations for the system directly. Comparing solutions from \textit{TAWAS} and \textit{Wigglewave} we see that our analytical solution is accurate within the limits of the WKB approximation but under-reports the wave damping, caused by enhanced phase mixing, beyond the WKB limit. Both \textit{TAWAS} and \textit{Wigglewave} solve the linearised governing equations and not the complete nonlinear MHD equations. Paper II will consider simulations that solve the full MHD equations including important nonlinear effects.
\end{abstract}

\begin{keywords}
(magnetohydrodynamics) MHD -- Plasmas -- Waves -- Sun: corona -- Sun: oscillations
\end{keywords}



\section{Introduction}
\label{sec:intro}

In this paper we will consider the propagation and enhanced phase mixing of torsional Alfv\'en waves in open structures in the solar corona with exponentially diverging magnetic field lines and a gravitationally stratified atmosphere. Wave based heating mechanisms have always been part of the discussion surrounding the coronal heating problem \citep{Arregui} and since the advent of space-based EUV imagers a large variety of waves and oscillations have been detected in the solar atmosphere \citep{Parnell}. In particular, torsional Alfv\'en waves have recently been directly detected in a magnetic pore at the photosphere \citep{Stangalini2021}. It was shown in \citep{Soler} that torsional Alfv\'en waves of intermediate frequencies are able to penetrate to the corona whilst low frequency waves are reflected and high frequency waves are absorbed in the chromosphere due to ion-neutral damping. Furthermore it was shown that the transmission of torsional Alfv\'en waves to the corona is improved in a magnetic field with exponentially diverging field lines.

Alfv\'en waves are believed to be viable transporters of the non-thermal energy required to heat the solar corona \citep{Mathioudakis} but they are notoriously difficult to dissipate due to the high conductivity of the corona, \citep{Jigsaw}. For this reason dissipation must occur over scales smaller than typical MHD scales. Mechanisms for generating these small length scales typically rely on inhomogeneities in the corona, \citep{DM2008}. Phase mixing was first proposed by \citep{Priest} as a mechanism for enhancing the viscous and ohmic dissipation of Alfv\'en waves in loops and open regions of the solar corona. When shear Alfv\'en waves on neighboring magnetic surfaces propagate at different speeds they move out of phase with each other. This leads to the build up of strong gradients perpendicular to the direction of propagation that allows dissipation of the wave through shear viscosity or resistivity heating the plasma. 


The efficiency of phase mixing is strongly affected by both the magnetic field configuration and the density stratification of the plasma, \citep{DM2009}. Enhanced phase mixing occurs in magnetic field structures with exponentially diverging field lines and significantly increases the dissipation rate of Alfv\'en waves. In this case the magnetic field strength decreases exponentially with height and therefore, so does the Alfv\'en speed. The effect is that as Alfv\'en waves propagate upwards, their wavelengths are decreased and this leads to stronger transverse gradients, which enhances the effect of phase mixing. Conversely density stratification causes wavelengths to increase as the waves propagate upwards and this reduces the effects of phase mixing \citep{Smith}.

In this paper we consider the effect of enhanced phase mixing on torsional Alfv\'en waves propagating upwards in an exponentially diverging field line structure with a transverse density gradient and stratified atmosphere. An analytic formulation for the evolution of shear Alfv\'en waves in a two-dimensional open magnetic field configuration is given by \citep{Ruderman1998} using the WKB approximation. In \citep{Smith} a corrected form of this analytical formulation was validated using a 2.5D visco-resistive linear MHD code and the dependency of the heating length scale on the magnetic scale heights was investigated. The conclusion was that the dissipation of Alfv\'en waves is strongly dependent on the field line divergence.

In both \citep{Ruderman1998} and \citep{Smith} it is assumed that the wavelength of Alfv\'en waves is much smaller than the characteristic scale of variation of the equilibrium parameters. This allows the WKB approximation to be used and the effect of wave reflection to be neglected. Wave reflection means that part of the Alfv\'en wave is reflected back along the same field line due to the longitudinal inhomogeneity in the Alfv\'en speed, as described, for example, in \citep{Morton}.

In \citep{Ruder2017} and \citep{Petru2018} non-reflective magnetic plasma configurations are considered; these are a planar and cylindrical configurations respectively. Exponentially divergent magnetic field lines are still used but the density profiles are altered to keep the configuration non-reflective. Phase mixed solutions for Alfv\'en waves propagating in these configurations are derived without using the WKB approximation. For these formulations the wavelength need not be much smaller than the characteristic scale of variation, it can be of the same order. The density profiles used however are only possible within a very restrictive range of parameters. To consider a more general configuration the effects of wave reflection must be included in the solutions. 

 In \citep{Ruderman2018} the WKB approximation is once again used to find the phase mixing solution of torsional Alfv\'en waves propagating in a cylindrically symmetric, potential magnetic field with exponentially diverging field lines, a radial density gradient and a stratified atmosphere. In this formulation the wavelength is not assumed to be much smaller than the characteristic scale of variation. Instead conditions are found in which the effect of wave reflection can be considered weak and neglected. 
 
 In this paper we begin by considering the analytic formulation for the propagation and phase mixing of torsional Alfv\'en waves. The next section \cref{sec:analytic_solutions} is split into three parts. In \cref{sec:analytic_formula} we go through the derivation of the analytic solution presented in \citep{Ruderman2018} making a simple correction towards the end. In \cref{sec:energy_calc} we describe how the wave energy flux of the Alfv\'en wave, $\Pi$, can be calculated over successive magnetic surfaces. Then in \cref{sec:setup} we describe the numerical setup for an equilibrium with divergent magnetic field lines and an exponentially stratified density profile which we use in our numerical calculations. 
 
 In \cref{sec:tawas} we present a numerical code, named \textit{TAWAS} and written in IDL, that calculates the analytical solution over a coordinate grid in radius, $r$ and height, $z$. We use \textit{TAWAS} firstly to demonstrate the effects of both wave reflection, and of our correction to the solution in \cref{sec:correction}. Then in \cref{sec:param_studies} we use \textit{TAWAS} to perform parameter studies to determine the conditions under which the damping of torsional Alfv\'en waves is strongest.
 
  In \cref{sec:wigglewave} we introduce a second code \textit{Wigglewave}, that directly solves the linearised governing equations for torsional Alf\'en wave propagation and is fourth order accurate in space and time. We compare the results from \textit{TAWAS} and \textit{Wigglewave} to test the validity of the analytical solution both within and beyond the limits of the WKB approximation. Our aim is to determine whether the wave damping caused by enhanced phase mixing is stronger or weaker than the analytic solution suggests. Finally in \cref{sec:conclusions} we discuss our results and summarise our findings.
  
  The overall aim of this work is to correct and validate the analytic solution presented in \citep{Ruderman2018} and to identify the conditions under which the WKB approximation, and by extension the analytic formula, is valid. In particular we will see whether Alfv\'en wave damping is stronger or weaker than predicted by this formula beyond the limits of the WKB approximation.

\section{Analytic Solutions}
\label{sec:analytic_solutions}
 
\subsection{Solution Formulation}
\label{sec:analytic_formula}
 
  We begin by considering the analytic solution for the propagation and phase-mixing of torsional Alfv\'en waves through a viscous plasma that is in axisymmetric equilibria. Such a solution is presented in \citep{Ruderman2018} and we will follow this solution closely, correcting a mistake in the solution towards the end of the derivation, as confirmed by \citep{Ruderman_private}.
  
  The only dissipative process considered here is viscosity whilst resistivity is neglected. In \citep{Priest} it is shown that the kinematic viscosity and magnetic diffusivity are additive in the dissipative term of the wave equation. As kinematic viscosity and magnetic diffusivity are of the same order in the corona neglecting resistivity can only reduce the efficiency of wave damping by a factor of unity; on the other hand this allows us to simplify the analysis by eliminating the magnetic perturbation from the governing equations.
  
\begin{figure}
\centering
  \includegraphics[width=0.7\linewidth]{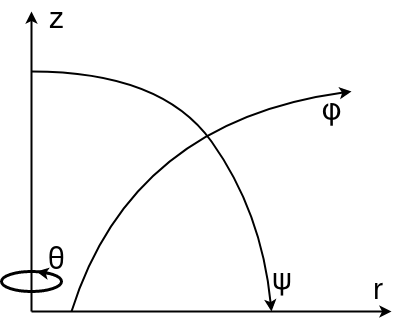}
\caption{A sketch of the cylindrical and curvilinear coordinate systems. In the cylindrical system we have the height, $z$ and the radius, $r$ whereas in the curvilinear field aligned coordinate system we have $\phi$, which varies along field lines and $\psi$, which varies across field lines. Both coordinate systems include the azimuthal coordinate $\theta$.}
\label{fig:sketch}
\end{figure}

  As our plasma is in axisymmetric equilibria all of the equilibrium quantities are dependent on only $r$ and $z$ in cylindrical coordinates. We consider an equilibrium magnetic field without azimuthal component $\mathbf{B_0} = (B_r,0,B_z)$. As the plasma beta in the corona is very low, we assume a negligible pressure. Our magnetic field must therefore be a force-free, axisymmetric magnetic field  without azimuthal component. This type of field is always potential and therefore we can express the radial and vertical components $B_r$ and $B_z$ in terms of the magnetic potential,  $\phi$, and magnetic flux, $\psi$,
 
\begin{equation}
\frac{B_r}{B_0} = \diffp{\phi}{r} = -\frac{H}{r}\diffp{\psi}{z}, \qquad
\frac{B_z}{B_0} = \diffp{\phi}{z} = \frac{H}{r}\diffp{\psi}{r},
\label{eq:coords}
\end{equation}
where $B_0$ is a constant characterizing the magnetic field strength and $H$ is the magnetic scale height which characterizes the divergence of the magnetic field lines with height. Below we use the curvilinear coordinate system $(\phi,\psi)$ in the plane $\theta =$ constant. By definition $\phi$ and $\psi$ are orthogonal with $\phi$ coordinate lines coinciding with the magnetic field lines and $\psi$ coordinate lines being orthogonal to the magnetic field lines. This coordinate system is sketched in \cref{fig:sketch}.

We begin our analysis with the MHD equations for a zero beta plasma,

\begin{align}
\label{eq:mhd}
    \diffp{\rho}{t} + (\mathbf{v}\cdot\nabla)\rho \quad = \quad &  -\rho(\nabla\cdot\mathbf{v}),
    \\
    \rho\left(\diffp{\mathbf{v}}{t} + (\mathbf{v}\cdot\nabla)\mathbf{v})\right)  \quad = \quad &
    -\frac{1}{\mu_0}\mathbf{B}\times(\nabla\times\mathbf{B}) + \nabla\cdot(\rho\nu\nabla\mathbf{v}),
    \\
    \diffp{\mathbf{B}}{t} + (\mathbf{v}\cdot\nabla)\mathbf{B})  \quad = \quad &
    (\mathbf{B}\cdot\nabla)\mathbf{v} -\mathbf{B}(\nabla\cdot\mathbf{v}),
\end{align}

We linearise these equations and consider purely azimuthal perturbations to the velocity and magnetic field of the plasma, so that only the $\theta$-components of the velocity and magnetic field perturbation, $v = v_\theta$ and $b = b_\theta$, are non-zero. This results in torsional Alfv\'en waves that are incompressible and are described by the $\theta$-components of the momentum and induction equation,

\begin{align}
\label{eq:velocity}
    \rho \diffp {v}{t}  = \quad &  
    \frac{1}{r\mu_{0}}\left(\mathbf{B_0}\cdot\nabla(rb)\right)
    + \frac{1}{r}\diffp{}{r}\left(\rho\nu r\diffp{v}{r}\right)
    + \diffp{}{z}\left(\rho\nu\diffp{v}{z}\right),
    \\
\label{eq:magnetic}
    \diffp {b}{t} = \quad &
    r\mathbf{B_0} \cdot\nabla\left(\frac{v}{r}\right),
\end{align}
where $\rho$ is the density, $\mu_0$ is the vacuum permeability, $\nu$ is the kinematic viscosity, which is taken to be constant, and $B_0\cdot\nabla = B_r\partial_r +B_z\partial_z$. We now make the substitution $v = ru$ where $u$ is the angular velocity. It is shown in \citep{Petru2018} that we can simplify the viscous terms in \cref{eq:velocity} by making the approximation,

\begin{equation}
\label{eq:visc_approx}
    \frac{1}{r}\diffp{}{r}\left(\rho\nu r\diffp{v}{r}\right)
    + \diffp{}{z}\left(\rho\nu\diffp{v}{z}\right) \; \approx \; 
    \rho\nu r \left(\diffp[2]{u}{r}+\diffp[2]{u}{z}\right),
\end{equation}
we can then take the time derivative of  \cref{eq:velocity} and then eliminate $b$ using \cref{eq:magnetic} to obtain the wave equation,

\begin{equation}
\diffp[2]{u}{t} \; = \; 
\frac{1}{\mu_0 \rho r^2}\mathbf{B_0}\cdot\nabla \left(r^2\mathbf{B_0}\cdot\nabla u\right)
+ \nu \left(\diffp{}{t}\diffp[2]{u}{r}+\diffp{}{t}\diffp[2]{u}{z}\right).
\end{equation}

We continue to follow \citep{Petru2018} and now switch to using the curvilinear magnetic field coordinates $\phi$ and $\psi$ as independent variables, our partial derivatives become,

\begin{alignat}{2}
\diffp{}{r} \quad =  \quad & 
\frac{B_r}{B_0}\diffp{}{\phi} +\frac{r}{H}\frac{B_z}{B_0}\diffp{}{\psi},
\\
\diffp{}{z} \quad = \quad &
\frac{B_z}{B_0}\diffp{}{\phi} -\frac{r}{H}\frac{B_r}{B_0}\diffp{}{\psi} ,
\\
\label{eq:bdotnabla}
\mathbf{B_0}\cdot\nabla \quad = \quad &
\frac{B^2}{B_0}\diffp{}{\phi},
\end{alignat}
where $B$ is the field strength of our equilibrium magnetic field $\mathbf{B_0}$. By assuming that derivatives in the $\psi$ direction, perpendicular to field lines, dominate over derivatives in the $\phi$ direction, along field lines, we can make the approximation,

\begin{align}
\diffp[2]{u}{r}+\diffp[2]{u}{z} \quad = \quad \frac{r^2 B^2}{H^2 B_0^2}\diffp[2]{u}{\psi},
\end{align}

\noindent which transforms our wave equation into:

\begin{align}
\label{eq:wave_eqn}
\diffp[2]{u}{t} \quad = \quad & \frac{V_A^2}{r^2} 
\diffp{}{\phi} \left(\frac{r^2B^2}{B_0^2}\diffp{u}{\phi} \right)+\frac{\nu r^2 B^2}{H^2 B_0^2}
\frac{\partial^3 u}{\partial t \partial \psi^2},
\end{align}

\noindent where 

\begin{align}
V_A \quad = \quad & \frac{B}{\sqrt{\mu_0 \rho}},
\end{align}

is the Alfv\'en velocity. We now look for a solution to \cref{eq:wave_eqn} of the form,

\begin{align}
\label{eq:u_decom}
u(t,\phi,\psi) \quad = \quad &  
\frac{v_0}{H}A(\phi,\psi)\Phi(t,h(\phi,\psi),\psi),
\end{align}
where $v_0$ is a constant of velocity. The functions $A(\phi,\psi)$ and $h(\phi,\psi)$ can be chosen arbitrary, we define them as:
 
\begin{align}
\label{eq:h_and_A}
h(\phi,\psi) \; = \; 
B_0V_0 \int_{\phi_1(\psi)}^{\phi} \frac{d\phi'}{BV_A},
\quad
A(\phi,\psi) \; = \; 
A_0(\psi)\frac{H}{r}\left(\frac{\rho_0}{\rho}\right)^{1/4},
\end{align}
 where $V_0 = B_0/\sqrt{\mu_0\rho_0}$, $\rho_0$ is the density at the origin and $\phi_1(\psi)$ $A_0(\psi)$ are arbitrary functions with the condition that $A_0(\psi)/r$ is nonzero in the limit $\psi \to 0$. With these functions defined, substituting \cref{eq:u_decom} into \cref{eq:wave_eqn} gives the following equation for $\Phi$:
 
\begin{align}
\label{eq:wave_Phi}
\diffp[2]{\Phi}{t} -
V_0^2\diffp[2]{\Phi}{h} \; = \; 
\frac{\nu r^2 B^2}{H^2 B_0^2}\diffp{\Xi}{t} +
\frac{V_A^2 \Phi}{A r^2}\diffp{}{\phi}\left(\frac{r^2B^2}{B_0^2}\diffp{A}{\phi} \right),
\end{align}
where

\begin{align}
\begin{split}
\Xi \; = \; &
\frac{1}{A}\diffp[2]{(A\Phi)}{\psi}
+ \frac{2}{A}\diffp{h}{\psi} \diffp{}{\psi}\left(A\diffp{\Phi}{h}\right)\\
& + \diffp[2]{h}{\psi}\diffp{\Phi}{\psi}
+ \left(\diffp{h}{\psi}\right)^2\diffp[2]{\Phi}{h}.
\end{split}
\end{align}

The terms on the RHS of \cref{eq:wave_Phi} prescribe the wave damping from viscous dissipation and wave reflection respectively. In order to apply the WKB approximation it is assumed that both the terms on the RHS of \cref{eq:wave_Phi} are much smaller than the terms on the LHS. We now assume Alfv\'en waves are driven at the lower boundary, representing the base of the corona, and then propagate along magnetic field lines. Considering harmonic waves and taking all time-dependant quantities as proportional to $e^{i\omega t}$ reduces \cref{eq:wave_Phi} to,

\begin{align}
\label{eq:wave_harmonic}
V_0^2\diffp[2]{\Phi}{h} +
\omega^2 \Phi \; = \; 
\frac{i\omega\nu r^2 B^2}{H^2 B_0^2}\Xi
-\frac{V_A^2 \Phi}{A r^2}\diffp{}{\phi}\left(\frac{r^2B^2}{B_0^2}\diffp{A}{\phi} \right).
\end{align}

Our assumption that wave damping is weak allows the use of the following scaled quantities in our wave equation, 

\begin{align}
\tilde{h}\; = \; \epsilon h, 
\qquad \tilde{\nu}\; = \; Re\;\nu, 
\qquad \tilde{\phi}\; = \epsilon \phi,
\end{align}
where the scaling parameter $\epsilon \ll 1$ and $Re = HV_0/\nu$ is the Reynolds number, see \citep{Ruderman2018} for details. Substituting these scaled quantities transforms \cref{eq:wave_harmonic} once again,

\begin{align}
\label{eq:wave_scaled}
\epsilon^2 V_0^2\frac{\partial^2 \Phi}{\partial\tilde{h}^2} +
\omega^2 \Phi \; = \; 
\frac{i\omega\tilde{\nu} r^2 B^2}{H^2 B_0^2 Re}\tilde{\Xi}
-\frac{\epsilon V_A^2 \Phi}{A r^2}\diffp{}{\tilde{\phi}}\left(\frac{r^2B^2}{B_0^2}\diffp{A}{\phi} \right).
\end{align}

At this stage the standard WKB method is used to look for a solution in the form,

\begin{align}
\label{eq:Phi_form}
\Phi \; = \; 
Q(\tilde{h},\psi)\exp{i\epsilon^{-1}\Theta(\tilde{h},\psi)},
\end{align}
where $\tilde{\Xi}$ is simply $\Xi$ but with $\tilde{h}$ substituted for $h$. Substituting this expression into \cref{eq:wave_scaled} and using the expression for $\tilde{\Xi}$ gives:

\begin{alignat}{2}
\label{eq:wave_WKB}
\begin{split}
\epsilon^2\frac{\partial^2 Q}{\partial \tilde{h}^2}
+ & 2i\epsilon\diffp{Q}{\tilde{h}}\diffp{\Theta}{\tilde{h}}
+ i\epsilon Q\frac{\partial^2 \Theta}{\partial \tilde{h}^2}
-  Q\left(\diffp{\Theta}{\tilde{h}}\right)^2
+ \frac{\omega^2}{V_0^2}Q \\
= & - \frac{i \epsilon^{-2}\omega \tilde{\nu} r^2 B^2}{H^2 B_0^2 V_0^2 Re}Q
\left[\left(\diffp{\tilde{h}}{\psi}\right)^2\left(\diffp{\Theta}{\tilde{h}}\right)^2 + \mathcal{O}(\epsilon)\right]\\
& - \frac{\epsilon V_A^2}{A r^2 V_0^2} Q\diffp{}{\tilde{\phi}}\left(\frac{r^2B^2}{B_0^2}\diffp{A}{\phi}\right).
\end{split}
\end{alignat}

Now terms of the same order are collected. Assuming that the first term on the RHS is smaller than the two largest terms on the LHS, terms of the order unity are collected. This is called the approximation of geometrical optics and determines the shape of the rays along which waves propagate. This gives us an equation for $\Theta$:

\begin{align}
\left(\diffp{\Theta}{\tilde{h}}\right)^2  \; = \; 
\frac{\omega^2}{V_0^2},
\end{align}

which when solved for waves propagating in the positive $\phi$ direction gives:

\begin{align}
\label{eq:Theta_equals}
\Theta \; = \; 
\frac{\tilde{h}\omega}{V_0}.
\end{align}

Now the terms of order $\epsilon$ are collected. This is called the approximation of physical optics and determines the spatial evolution of the wave amplitude. The first term on the RHS which describes resistive damping due to phase mixing should be included in this contribution hence we set $Re = \epsilon^{-3}$.

This is where our correction is made and our solution diverges from that given in \citep{Ruderman2018}. In eq.(18) of \citep{Ruderman2018} when the terms of order $\epsilon$ are collected all the terms are divide through $i$ except for the second term on the RHS. Here we correct this by dividing all terms by $i$. This gives us,

\begin{alignat}{2}
\begin{split}
2\diffp{Q}{\tilde{h}}\diffp{\Theta}{\tilde{h}}
+ Q\frac{\partial^2 \Theta}{\partial \tilde{h}^2}
=  - \frac{\omega \tilde{\nu} r^2 B^2}{H^2 B_0^2 V_0^2}Q
\left(\diffp{\tilde{h}}{\psi}\right)^2\left(\diffp{\Theta}{\tilde{h}}\right)^2 \\
+ \frac{i V_A^2}{A r^2 V_0^2} Q\diffp{}{\tilde{\phi}}\left(\frac{r^2B^2}{B_0^2}\diffp{A}{\phi}\right). \end{split}
\end{alignat}

Returning to non-scaled variables now gives the following ordinary differential equation for Q:

\begin{align}
\label{eq:damping_Q}
\frac{\partial Q}{\partial \phi}
\quad = \quad
\left(
iR- \Upsilon
\right) Q,
\end{align}
where,

\begin{align}
\label{eq:Upsilon}
\Upsilon \quad = & \quad \frac{\omega^2 \nu}{2 V_0^3 G^2}
\left(\frac{\partial h}{\partial \psi}\right)^2, \\
\label{eq:Reflection}
R \quad = & \quad \frac{G V_0}{2 \omega H^2} \diffp{}{\phi}\left(\frac{r^2B^2}{B_0^2}\diffp{}{\phi} G\right),
\end{align}

and $G=A/A_0$. We also impose the condition:

\begin{align}
Q \; = \; 1 \quad \text{at} \quad \phi = \phi_1(\psi).
\end{align}

As with \citep{Ruderman2018} we can now use \cref{eq:Theta_equals,eq:Phi_form,eq:h_and_A,eq:u_decom} and recall that the perturbations are proportional to $e^{-i\omega t}$ to obtain:

\begin{align}
\label{eq:v_soln}
v \; = \; v_0 W \exp{[i\omega(h/V_0 - t)]}, \qquad 
W = QA_0(\psi)\left(\frac{\rho_0}{\rho}\right)^{1/4}.
\end{align}

For the magnetic field perturbation we can use \cref{eq:magnetic,eq:h_and_A,eq:v_soln} and the relation \cref{eq:bdotnabla} to yield,

\begin{align}
\label{eq:b_soln}
b \; = \; v_0 B \left[\frac{irB}{\omega B_0}\diffp{(W/r)}{\phi}
-\frac{W}{V_A}\right]\exp{[i\omega(h/V_0-t)]}.
\end{align}

These solutions differ from those given in \citep{Ruderman2018} as $Q$ and consequently $W$ now have different values, furthermore they are now complex valued \citep{Ruderman_private}. 

\subsection{Wave Energy Flux Calculation}
\label{sec:energy_calc}

In order to characterise the efficiency of wave damping we follow the example in \citep{Ruderman2018} being careful to account for the fact that $Q$ and $W$ are now complex valued. We calculate the wave energy flux across magnetic surfaces defined by $\phi =$ constant. Each surface is uniquely identified by its height of intersection with the $z$-axis, hence we can define the energy flux $\Pi$ as a function of height $z$. Multiplying \cref{eq:velocity} by $v$, \cref{eq:magnetic} by $b/\mu_0$ and adding the results we obtain,

\begin{align}
\begin{split}
\label{eq:wave_energy}
\frac{\partial}{\partial t} \left(\frac{\rho v^2}{2} + \frac{b^2}{2\mu_0} \right) \; = \;& 
\frac{1}{\mu_0} \nabla\cdot(\mathbf{B_0}vb) \\ &
+\frac{v}{r}\frac{\partial}{\partial r}
\left(\rho\nu r \diffp{v}{r}\right) +
v\frac{\partial}{\partial z} \left(\rho\nu\diffp{v}{z}\right).
\end{split}
\end{align}

The expression in the parenthesis on the LHS is the wave energy density, whilst $\mathbf{B_0}vb/\mu_0$ is the density of wave energy flux. When $\nu=0$ this is the equation for wave energy conservation. The density of wave energy flux is directed along the field lines. We now want to obtain an expression for the density of wave energy flux averaged over a wave period. 

To obtain physical quantities we need to take the real parts of $v$ and $b$ as given in \cref{eq:v_soln} and \cref{eq:b_soln}. It is however no longer possible to reduce this quantity to a simple expression in terms of $W$, for example by evaluating $\partial (W/r)/\partial\phi$ we see that the expansion of $b$ is,

\begin{align}
\begin{split}
\label{eq:b_expanded}
b \; = \;& 
-v_0 W \left[ \frac{iB^2}{\omega B_0}\left[
\frac{1}{Q}\diffp{Q}{\phi}+\frac{1}{4\rho}\diffp{\rho}{\phi}+\frac{1}{r}\diffp{r}{\phi}\right] + \sqrt{\mu_0\rho}\right] \\
    & \qquad\exp{[i\omega(h/V_0-t)]}.
\end{split}
\end{align}

Simultaneously we are faced with the issue that in order to evaluate the density of wave energy flux for our simulation results we will have to use the values of $v$ and $b$ directly. For both of these reasons we decide to calculate wave energy flux through each surface using the wave envelopes for $v$ and $b$. By using the assumption that $v$ and $b$ are in anti-phase we can conclude that the density of wave energy flux averaged over a wave period is,

\begin{align}
\label{eq:density_vp_bp}
-\frac{B}{\mu_0}\langle vb \rangle = \frac{B}{2 \mu_0} v_p b_p,
\end{align}
where $v_p$ and $b_p$ are the maximum values of the wave envelopes for $v$ and $b$ respectively. We now define $\Sigma$ as the magnetic surface defined by $\phi_* =$ constant, that intersect the $z$-axis at $z_*$. We assume that the waves propagate in the magnetic tube bounded by the surface $\psi = \psi_b$. The wave energy flux through $\Sigma$ is then the density of wave energy flux integrated over the part of $\Sigma$ in the magnetic tube, we denote this part as $\Sigma_b$. It is shown in \citep{Ruderman2018} that,

\begin{align}
\label{eq:dsigma}
d\Sigma = \frac{HB_0}{B} d\psi d\theta.
\end{align}

Now using \cref{eq:density_vp_bp} and \cref{eq:dsigma} we obtain that the average energy flux through the surface $\Sigma_b$ is,

\begin{align}
\label{eq:energy}
\Pi(z) \; = \; 
-\frac{1}{\mu_0}\int_{\Sigma_b}{B\langle vb \rangle} \; d\Sigma \; = \;
\pi \frac{HB_0}{\mu_0} \int_{0}^{\psi_b} v_p b_p \; d\psi.
\end{align}

\subsection{Numerical Setup}
\label{sec:setup}
 
We can now use our analytic solution to solve for the propagation of torsional Alfv\'en waves in a potential magnetic field with exponentially diverging field lines and a stratified density profile. First we must define the magnetic structure and density profile, our setup is identical to that given in \citep{Ruderman2018}.

We consider a magnetic field that is potential, axisymmetric and has no azimuthal component. The field must therefore satisfy the vector Laplace equation $\nabla^2\mathbf{B_0}=0$ in cylindrical coordinates, the solution to this equation is outlined in \cref{App1}. Solving this equation with the criteria that the field strength decreases with height exponentially according to our magnetic scale height $H$, defines our magnetic field as,

\begin{align}
B_r = B_0e^{-z/H}J_1(r/H), \qquad 
B_z = B_0e^{-z/H}J_0(r/H),
\end{align}
where $J_0$ and $J_1$ are Bessel functions of the first kind and of zero and first order respectively. The functions $\phi$ and $\psi$ are then given by,

\begin{align}
\phi = -He^{-z/H}J_0(r/H), \qquad 
\psi = re^{-z/H}J_1(r/H).
\label{eq:define_magcoord}
\end{align}

We define our gravitationally stratified atmosphere as follows,

\begin{align}
\rho \; = \;  \hat{\rho}(\psi)e^{-z/H_\rho}
 \; = \; \hat{\rho}(\psi)e^{-\alpha z/H},
\end{align}
where the density scale height $H_\rho = k_BT_0/mg$, $k_B$ is Boltzmann's constant, $T_0$ is the temperature in the corona and is constant, $m \approx 0.6 m_p$ in the corona, $m_p$ is the proton mass, $g$ is the solar surface gravity,  $\alpha = H/H_\rho$ and $\hat{\rho}(\psi)$ is an arbitrary function that defines the density variation across magnetic field lines. 

\begin{figure}
  \includegraphics[width=\linewidth]{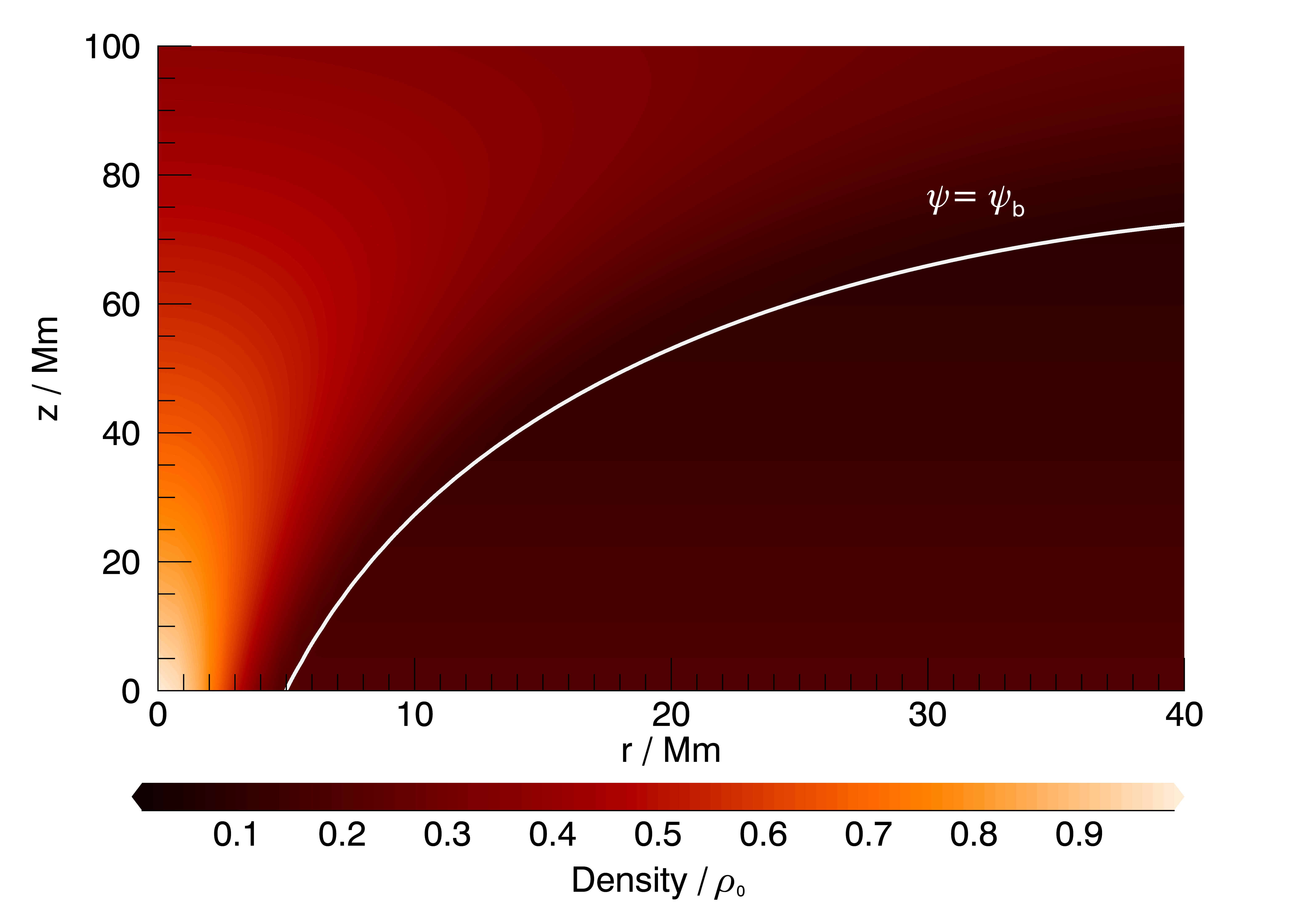}
\caption{An example of the density contrast for this setup, brighter areas represent a higher density, the white contour shows the boundary of the central high density tube. For this particular example the scale height used are $H = 20$ Mm and $H_\rho = 100$ Mm.}
\label{fig:field_line}
\end{figure}

We use $\hat{\rho}(\psi)$ to define a central higher density tube enclosed within the field line defined by $\psi = \psi_b$, we will study the propagation of torsional Alfv\'en waves within this tube. We want this tube to have a density gradient transverse to the propagation of Alfv\'en waves along magnetic field lines so that we can study the effect of phase mixing. We do this by defining $\hat{\rho}(\psi)$ to be,

\begin{align}
\hat{\rho}(\psi) = \frac{\rho_0}{\zeta}
\begin{cases}
    1+(\zeta-1)(1-\psi/\psi_b)^2,& \psi \leq \psi_b\\
    1,              & \psi \geq \psi_b
\end{cases}
\end{align}
where $\zeta$ is the density contrast between the centre and exterior of the magnetic tube $\hat{\rho}(0)/\hat{\rho}(\psi_b)$. An example of what the density structuring looks like is given in \cref{fig:field_line} the tube boundary at $\psi = \psi_b$ is highlighted for clarity.

Finally we define our wave driving, we would like to drive our waves from the lower boundary of the domain however strictly speaking we should be driving waves from a magnetic surface defined by $\phi = const$, let us consider the magnetic surface through the coordinate origin $r=0$,$z=0$ which has a value $\phi = -H$. Let this surface intersect $\psi = \psi_b$ at $r = r_0$ and $z = z_0$, from \cref{eq:define_magcoord} we then have,

\begin{align}
\psi_b \; = \; r_0\frac{J_1(r_0/H)}{J_0(r_0/H)}, \qquad
z_0 \; = \;  -Hln(J_0(r_0/H)).
\end{align}

By taking $r_0 \ll H$ we can approximate the Bessel functions by the leading terms in their Taylor series, $J_0(r_0/H) \approx 1$, $J_1(r_0/H) \approx r_0/2H$. This leads us to,

\begin{align}
\psi_b \; \approx \; r_0^2/2H, \qquad
z_0 \; \approx \;  0.
\label{eq:psib_approx}
\end{align}

This implies that the magnetic surface $\phi = -H$ roughly coincides with the lower boundary $z = 0$ within the confines of the tube meaning we can treat the lower boundary as a magnetic surface. The radius $r_0$ is where the tube boundary $\psi = \psi_b$ intersects the lower boundary and defines the width of the tube. We can therefore drive our Alfv\'en waves from the lower boundary $z = 0$ and within the tube structure $r \leq r_0$, we define our wave driving by,

\begin{align}
u\; = \; 
\begin{cases}
    u_0\left(1 -\frac{r^2}{r_0^2}\right)e^{-i\omega t},    & r \leq r_0\\
    0,  & r \geq r_0
\end{cases}
\qquad \text{at} \quad z = 0
\end{align}

From this expression for the angular velocity $u$ we can find the linear velocity $v = ru$ which we will use when driving waves in our simulations. We can calculate $A_0(\psi)$ which is required for the analytic solution. By comparing this expression for $u$ with \cref{eq:v_soln}, noticing that $h=0$ at $z=0$ and using the relation $r^2/r_0^2 = \psi/\psi_b$, we have,

\begin{align}
A_0(\psi)\; = \; 
\begin{cases}
    \left(\frac{\hat{\rho}}{\rho}\right)^{1/4} \left(\frac{\psi}{\psi_b}\right)^{1/2}
    \left(1-\frac{\psi}{\psi_b}\right),    & \psi \leq \psi_b \\
    0,  & \psi \geq \psi_b
\end{cases}
\end{align}

In what follows we consider the propagation of Alfv\'en waves in a typical coronal plume or divergent coronal loop structures. We set our characteristic values for field strength and density as $B_0 = 0.001$ T, and $\rho_0 = 1.66 \times 10^{-12}$ kg m\textsuperscript{-3} (which corresponds to a number density of approximately $n_0 = 1\times10^{15}$ m\textsuperscript{-3}). This gives us an Alfv\'en speed of $V_0 \approx 700$ m s\textsuperscript{-1}. We fix the initial tube radius to be $r_0 = 5$ Mm and the density contrast to be $\zeta = 5$ for all cases. We set the amplitude of our Alfv\'en wave driving at the lower boundary, representing the base of the corona, to have a maximum value of about 40 km s\textsuperscript{-1}, based on \citep{Banerjee} and \citep{Doyle}, this is equivalent to setting $u_0 = 100$ km s\textsuperscript{-1}.

The parameters that we vary across different cases are the magnetic and density scale heights, $H$ and $H_\rho$, the wave period $T$ and the kinematic viscosity $\nu$. The magnetic and density scale heights used are between 20 and 100 Mm. The magnetic scale height depends very much on the structure being considered, however, considering the temperature of the corona is 1-2 MK, see \citep{Aschwanden}, we would expect the density scale height be somewhere between 50-100 Mm. The wave periods considered are between 10 and 120 seconds as we expect shorter period waves to be absorbed in the chromosphere and longer period waves to be reflected as discussed in \citep{Soler}. 
The values we use for kinematic viscosity range between $\nu = 1\times 10^4$ and $5\times 10^7$ m\textsuperscript{2}s\textsuperscript{-1}, these values are several orders of magnitude greater than would be expected from classical plasma theory \citep{Braginskii}. This anomalous viscosity could be caused by turbulence as described in \citep{Liu}, possibly triggered by the Kelvin-Helmholtz instability as discussed in \citep{Resonance}, indeed numerical evidence for such values is presented in \citep{Tsik_anomalous}. 

\section{Analytic Solver: TAWAS}
\label{sec:tawas}

Using this numerical setup we proceed to calculate solutions for torsional Alfv\'en wave propagation for different cases, with each case using different values for the magnetic and density scale heights, $H$ and $H_\rho$, the wave period $T$ and the kinematic viscosity $\nu$. We calculate solutions using our analytical formula and an IDL script written by the authors called \textit{TAWAS (Torsional Alfv\'en wave analytic solution)}.

\textit{TAWAS} calculates the analytic solutions for $v$ and $b$ over a uniform numerical grid using the analytic formula described in \cref{sec:analytic_formula} and then calculates the wave energy flux $\Pi(z)$ as a function of the height using \cref{eq:energy} in \cref{sec:energy_calc}. The script for this code can be found at \url{https://github.com/calboo/TAWAS} under the filename \textit{tawas.pro}, along with a description of how the calculations are performed. 

Every time \textit{TAWAS} was used in this study the resolution was set to $500\times2000$ in the $r$ and $z$ directions respectively. The domain height was fixed at $z_{\text{max}} = 100$ Mm but the domain radius $r_{\text{max}}$ was varied depending on $H$ to allow for maximal resolution in each case. 

\begin{figure}
  \includegraphics[width=\linewidth]{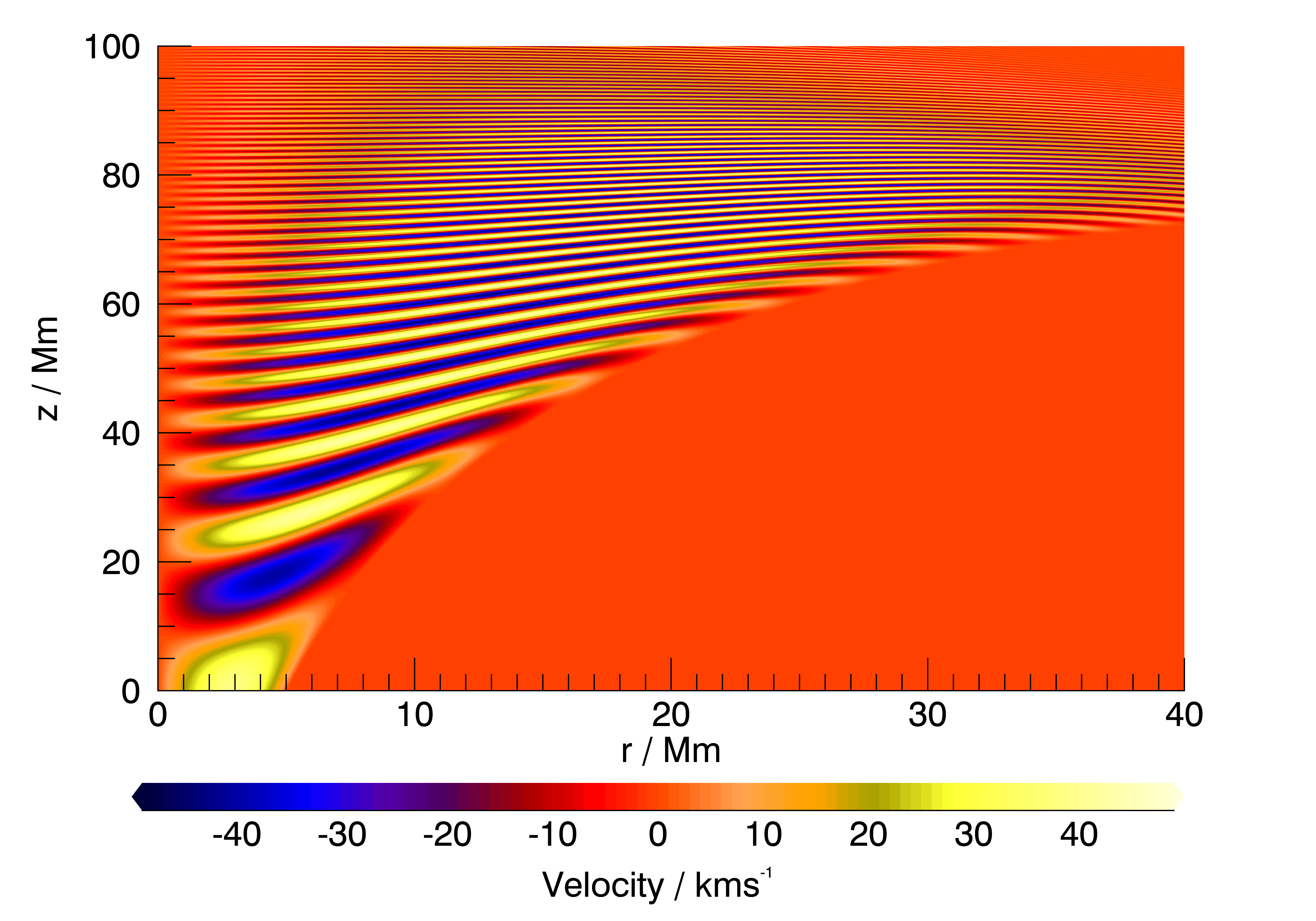}
\caption{An example of the velocity output from \textit{TAWAS}. This contour plot shows the azimuthal velocity $v$ of the plasma as the torsional Alfv\'en wave propagates through the central tube. Note that at higher heights the wavelength become very small so the wave fronts can only be resolved fully when zoomed in.}
\label{fig:v_example}
\end{figure}

\begin{figure}
  \includegraphics[width=\linewidth]{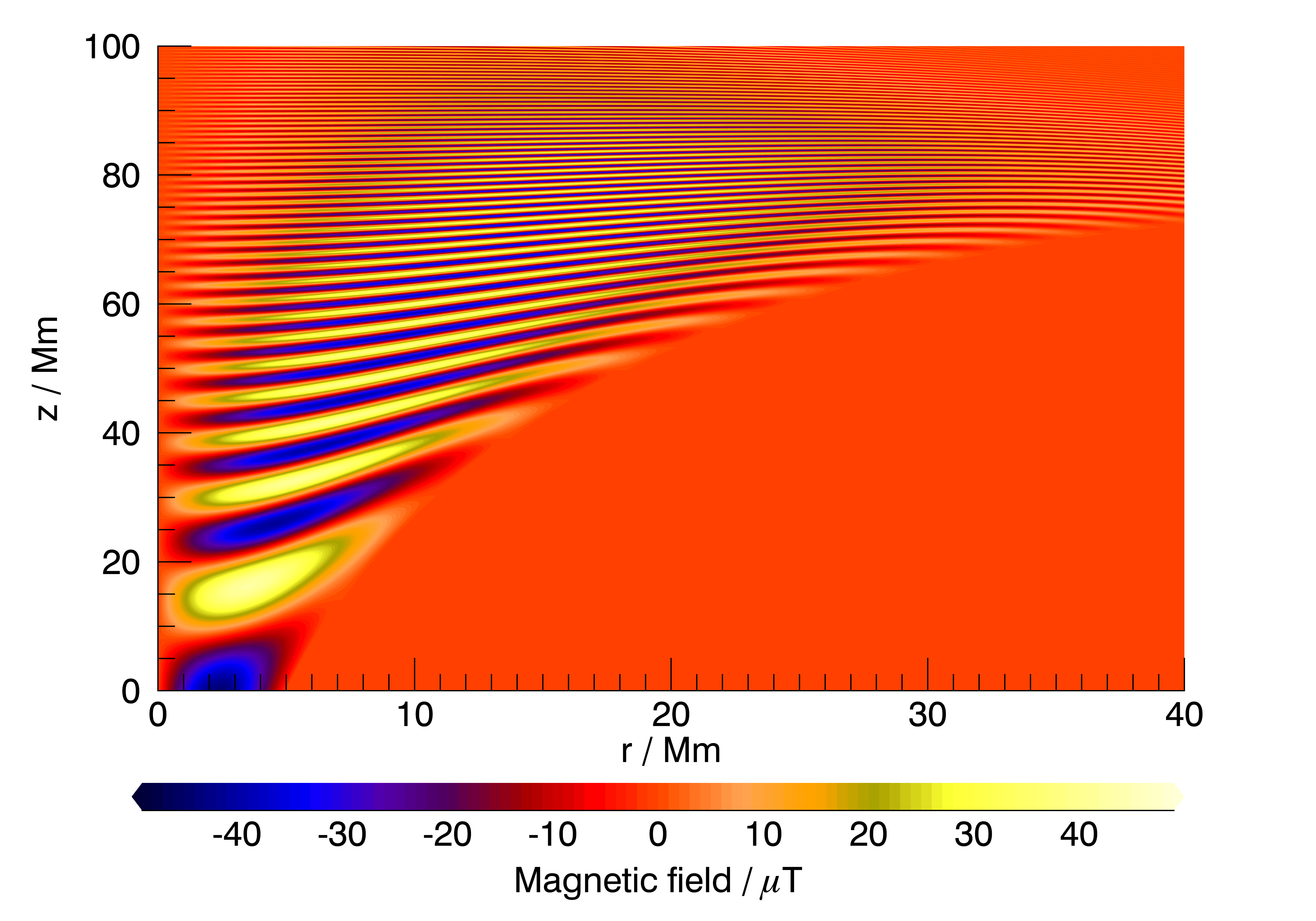}
\caption{An example of the magnetic field output from \textit{TAWAS}. This contour plot shows the azimuthal magnetic field $b$ of the plasma as the torsional Alfv\'en wave propagates through the central tube. Note that at higher heights the wavelength become very small so the wave fronts can only be resolved fully when zoomed in.}
\label{fig:b_example}
\end{figure}

\begin{figure}
  \includegraphics[width=\linewidth]{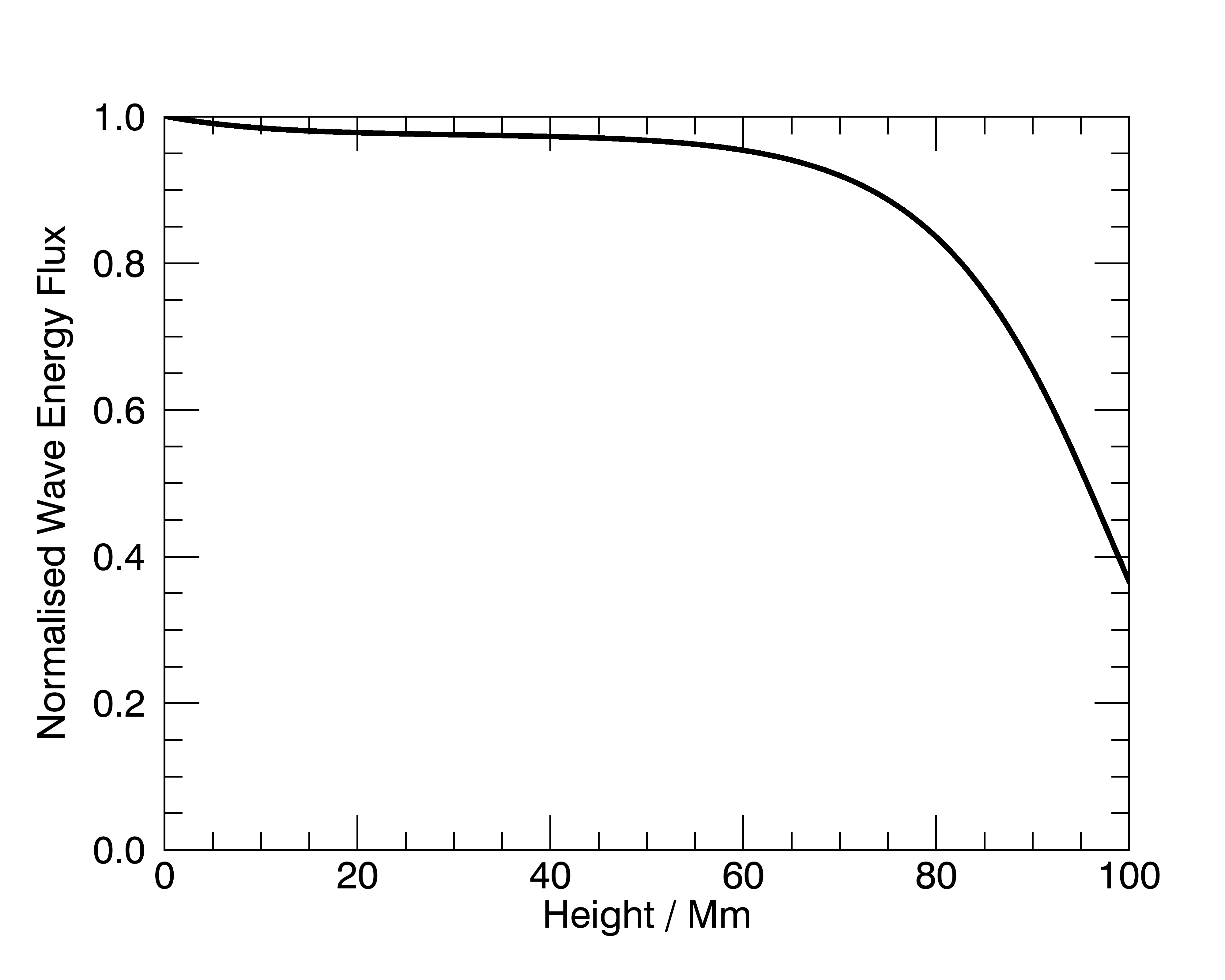}
\caption{An example of wave energy flux output from \textit{TAWAS}. This plot shows the normalised wave energy flux of the torsional Alfv\'en wave as it propagates upwards through the central tube.}
\label{fig:energy_example}
\end{figure}

Examples of the output from \textit{TAWAS} are shown in \cref{fig:v_example,fig:b_example,fig:energy_example}, these were produced from a case with the parameters $H = 20$ Mm, $H_\rho = 100$ Mm, $T = 60$ s and $\nu = 5 \times 10^7$  m\textsuperscript{2}s\textsuperscript{-1}. \cref{fig:v_example} and \cref{fig:b_example} show the velocity and magnetic field evolution of the wave over the domain and \cref{fig:energy_example}  shows the normalised wave energy flux, $\Pi(z)$, plotted against height. 

We can see the damping effects of both reflection and phase mixing in \cref{fig:energy_example}. Reflective damping occurs lower in the domain and its effect decreases with height whereas the damping caused by phase mixing is initially small and its effect increases with height. This is because the reflection and phase mixing of Alfv\'en waves depends on their wavelength. Longer wavelengths are reflected more strongly and shorter wavelengths undergo more phase mixing due to stronger transverse gradients that lead to greater viscous dissipation \citep{Smith}. In this case the divergence of the magnetic field lines causes the Alfv\'en velocity and therefore the wavelength to decrease with height thus reducing the effect of reflection and enhancing the effect of phase mixing as height increases. 

\section{Effect of correction to formula}
\label{sec:correction}
 
Now that we have a tool for numerically solving the analytic formula, we want to show the effect of our correction to the formula. Our correction only affects damping through wave reflection. In \citep{Ruderman2018} the effect of reflection was neglected entirely in numerical calculations. However, as we will see, the reflective term can cause significant damping to the total wave energy flux.

Here we will demonstrate two separate points. Firstly, that the effect of reflection causes wave damping that is significantly different from the case where reflection is neglected entirely. To do this we will need to compare the solutions from \textit{TAWAS} to those from an altered version of \textit{TAWAS} that does not include wave reflection. We have named this altered version \textit{norefl} and it can also be found at \url{https://github.com/calboo/TAWAS} under the filename \textit{norefl.pro}.

Secondly, that the effect of our correction to the formula caused wave damping that is significantly different from the uncorrected formula in which reflection is included. To do this will need to compare the solutions from \textit{TAWAS} to those from an altered version of \textit{TAWAS} that includes wave reflection but does not include our correction to the formula. We have named this altered version \textit{uncorr} and it can also be found at \url{https://github.com/calboo/TAWAS} under the filename \textit{uncorr.pro}.

To compare the wave energy flux outputs from these three scripts (\textit{TAWAS}, \textit{norefl} and \textit{uncorr}) we consider plots of $\Pi(z)$ from each script for two different cases. For both cases we set the magnetic and density scale heights to $H = 20$ Mm and $H_\rho = 100$ Mm respectively and the viscosity to $\nu = 5 \times 10^7$ m\textsuperscript{2}s\textsuperscript{-1}, this will allow us to see the effects of both wave reflection and of viscous dissipation.

\begin{figure}
  \includegraphics[width=\linewidth]{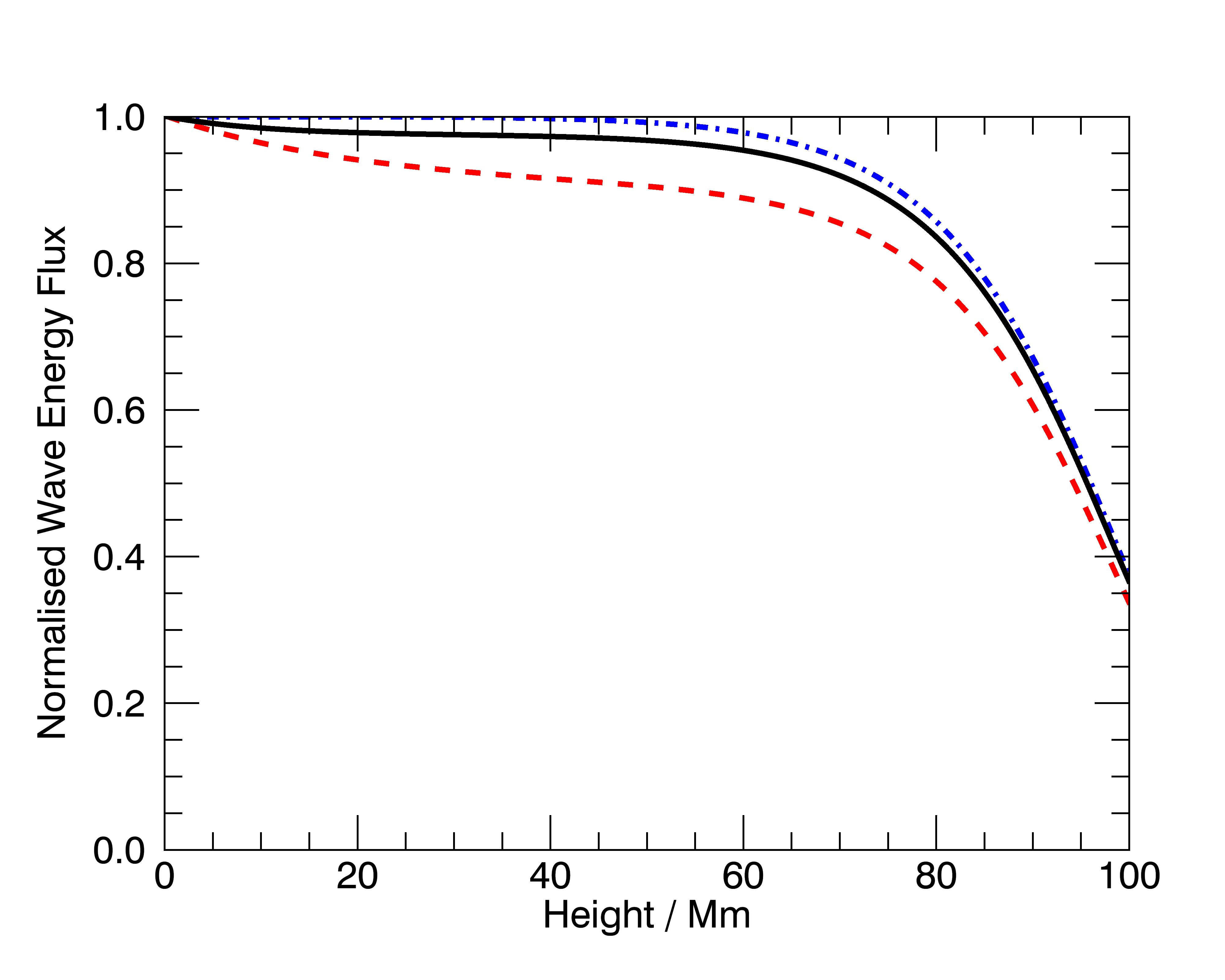}
\caption{Graph of normalised wave energy flux $\Pi(z)$ for the case in which $H = 20$ Mm, $H_\rho = 100$ Mm, $\nu = 5\times10^7$ m\textsuperscript{2}s\textsuperscript{-1} and $T = 60$ s. The dot-dashed blue line show results from the formula in which reflection is neglected, the dashed red line shows results from the formula that includes reflection but does not include our correction and the black line shows results from our corrected formula.}
\label{fig:correction_case2}
\end{figure}

\begin{figure}
  \includegraphics[width=\linewidth]{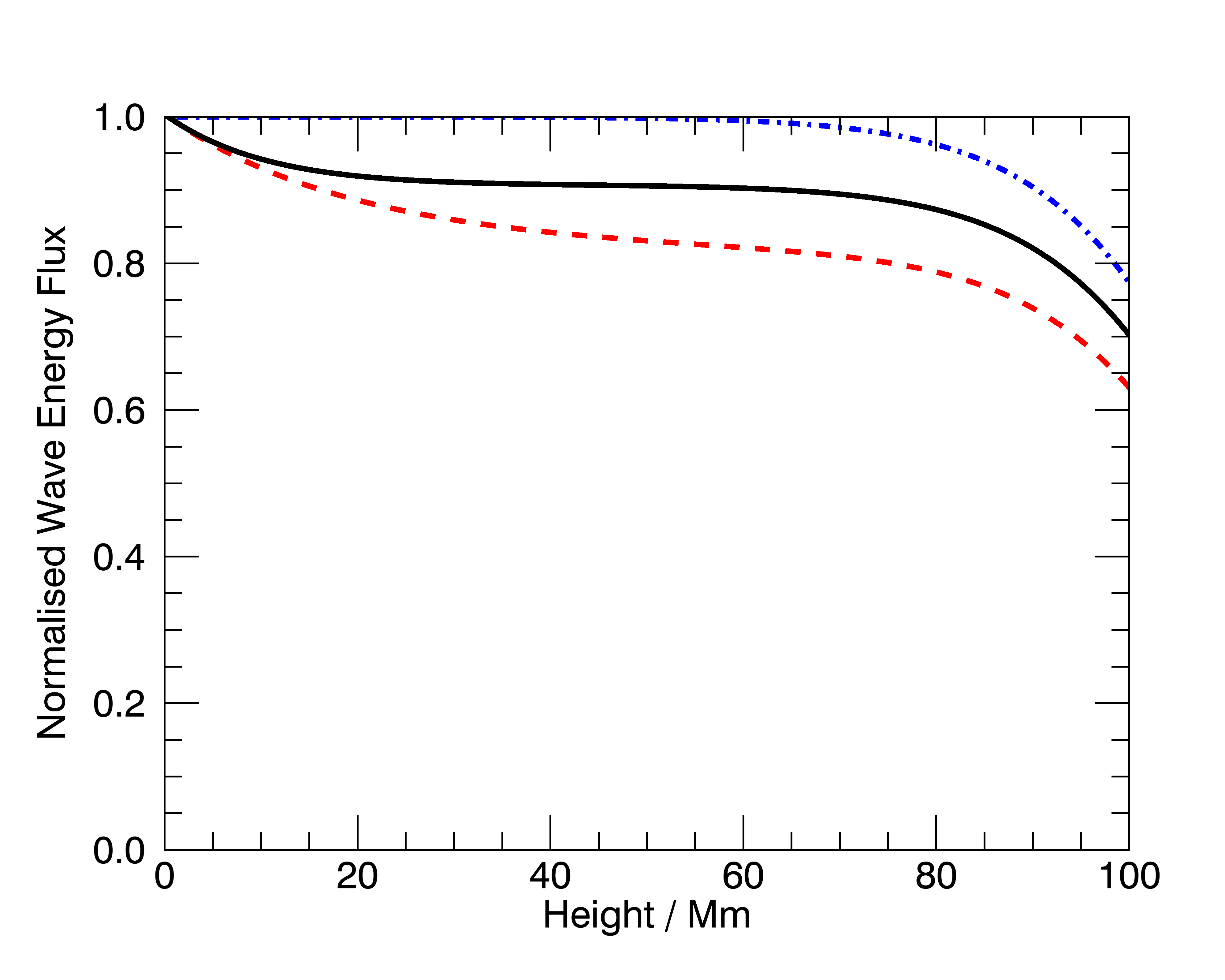}
\caption{Graph of normalised wave energy flux $\Pi(z)$ for the case in which $H = 20$ Mm, $H_\rho = 100$ Mm, $\nu = 5\times10^7$ m\textsuperscript{2}s\textsuperscript{-1} and $T = 120$ s. The dot-dashed blue line show results from the formula in which reflection is neglected, the dashed red line shows results from the formula that includes reflection but does not include our correction and the black line shows results from our corrected formula.}
\label{fig:correction_case3}
\end{figure}

For the first case we set the wave period to $T=60$ s. A graph of $\Pi(z)$ is shown for this case in \cref{fig:correction_case2}. For the second case we set the wave period to $T = 120$ s increasing the effect of wave reflection and reducing the effect of phase mixing. A graph of $\Pi(z)$ is shown for this case in \cref{fig:correction_case3}. 

We can see from \cref{fig:correction_case2} that when $T = 60$ s the effect of wave reflection would seem large if the uncorrected formula is used, however, once the formula is corrected the solution including reflection is very similar to the solution without wave reflection. In \cref{fig:correction_case3}, on the other hand, we can see that when $T = 120$ s the effect of wave reflection is still significant even when the corrected formula is used. 

From these graphs then, we can see that the effect of wave reflection should be included in our analysis as in some cases it can have a significant impact on wave damping. We can also see that our correction to the formula significantly changes the predicted amount of wave damping when compared to results from the uncorrected formula.

It is important when considering these graphs to note that, only
the damping caused by phase mixing will heat up the corona through viscous dissipation. It is therefore incorrect to equate all of the damping of wave energy flux to viscous heating power. At first glance this might suggest that neglecting wave reflection is a sensible idea, however, note that in \cref{fig:correction_case2} that the wave energy flux for both our corrected formula and for the solution with no reflection converge at higher heights. This suggests that if wave reflection is neglected then more power will be erroneously attributed to viscous heating. 

These graphs not only demonstrate the significance of our correction to the formula they also demonstrate the importance of calculating damping due to wave reflection. Wave spectra of Alfv\'en waves in the corona suggest that partial reflection of the waves occurs and is a frequency dependent phenomena \citep{Morton} as with our analytical solutions. By explicitly calculating the effect of wave reflection we can say with more certainty in which cases it can truly be neglected.

\section{Analytical Parameter Studies}
\label{sec:param_studies}

Using our corrected analytical formula and our numerical solver \textit{TAWAS} we can now find solutions for a variety of different cases. We now conduct parameter studies to identify the cases in which damping of the torsional Alfv\'en wave is strongest. 

This is necessary for two reasons. Firstly, we would like to know the conditions in which we expect enhanced phase mixing to dissipate a significant portion of the Alfv\'en waves energy through viscous dissipation. This will help to identify the conditions and regions of the solar corona in which phase mixing can fulfil the coronal heating requirement.

Secondly, we would like to test the validity of our analytic formula in the limit of strong Alfv\'en wave damping. When wave damping, in the form of either reflection or phase mixing, occurs over length scales comparable to the wavelength of the propagating wave then the WKB approximation is no longer valid, we say that these cases are beyond the WKB limit. 

By identifying the cases for which wave damping is strongest we can later test the validity of our analytic formula beyond the WKB limit by comparing results from \textit{TAWAS} to results from codes which solve the governing equations directly using finite differencing methods for these cases. Although we do not expect our formula to be accurate beyond the WKB limit, it is not clear whether we should expect more or less damping than predicted in these cases.

\subsection{Magnetic and Density Scale Height Parameter Study}
\label{sec:scale_param}
 
To begin with we perform a parameter study to investigate the effect of the magnetic and density scale heights $H$ and $H_\rho$. We run \textit{TAWAS} for values of $H$ and $H_\rho$ between 20 and 100 Mm, viscosity and wave period are fixed at $\nu = 5 \times 10^7$ m\textsuperscript{2}s\textsuperscript{-1} and $T = 60$ s. We very broadly want to measure the total wave damping for each case, to quantify this we consider the normalised wave energy flux remaining at height $z_{\text{max}}$, that is $\Pi(z_{\text{max}})/\Pi(0)$.

\begin{figure}
  \includegraphics[width=\linewidth]{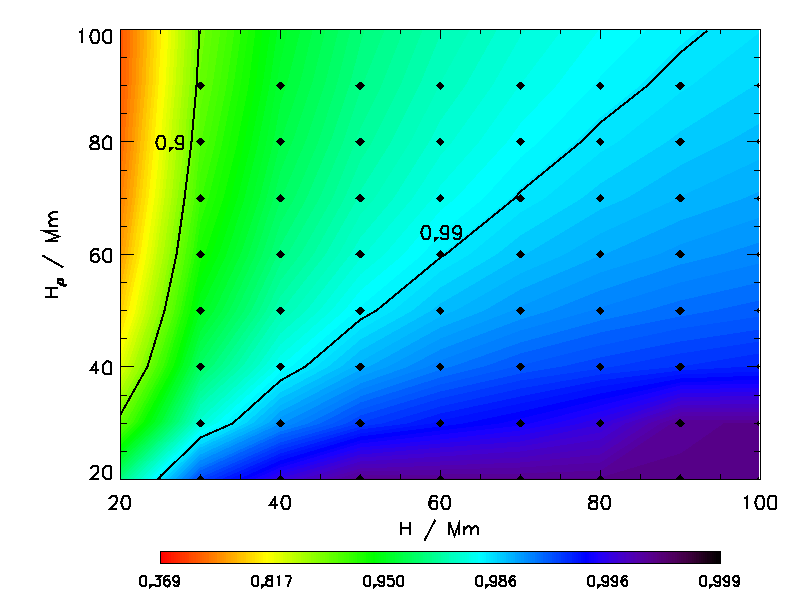}
\caption{Contour plot showing values of the normalised wave energy flux, $\Pi(z_{\text{max}})/\Pi(0)$, from our scale height parameter study. Note that the colour scale is logarithmic. The magnetic scale height $H$ is on the $x$-axis and the density scale height $H_\rho$ is on the $y$-axis. The diamond symbols represent data points from the parameter study and the thick line contours indicate where $\Pi(z_{\text{max}})/\Pi(0)$ has values of 0.9 and 0.99.}
\label{fig:scale_grid}
\end{figure}

In \cref{fig:scale_grid} we see a contour plot of $\Pi(z_{\text{max}})/\Pi(0)$ for the different values of $H$ and $H_\rho$ used in this study. We can see that for $\nu = 5 \times 10^7$ m\textsuperscript{2}s\textsuperscript{-1} and $T = 60$ s there is only significant damping when the magnetic field lines are highly divergent. In fact the only cases where $\Pi(z_{\text{max}})/\Pi(0) < 0.9$ were those with $H = 20$ Mm. Furthermore we can see that the damping increases as the density scale height is increased. 

These results are what one would expect. The mechanism of enhanced phase mixing is greater for equilibria with highly divergent field lines and weak density stratification. The divergence of the magnetic field lines causes a decrease in Alfv\'en speed with height, this in turn decreases the wavelength of propagating waves leading to stronger transverse gradients and increased phase mixing. Conversely atmospheric stratification increases the Alfv\'en speed with height, increasing the wavelength of waves and reducing the effect of phase mixing.

By looking at graphs of $\Pi(z)$, such as the graph shown in \cref{fig:energy_example}, for each individual case we can also see that wave reflection is strongest for low values of $H$ and high values of $H_\rho$. This is however for a different reason. Wave reflection is stronger when Alfv\'en waves encounter strong longitudinal gradients in the Alfv\'en speed. In the case of highly divergent magnetic field lines and weak density stratification, the gradient in Alfv\'en speed along field lines, i.e. in the direction of wave propagation, is greatest and hence the effect of wave reflection is strongest. 

\subsection{Viscosity-Period Parameter Study}
\label{sec:visc-period}

Having identified the scale heights for which wave damping is the strongest we move on to investigate the effects of changing the viscosity, $\nu$, and the wave period, $T$. We conduct a second parameter study using \textit{TAWAS}. We vary $\nu$ between $1\times10^{-4}$ and $\nu = 5 \times 10^7$ m\textsuperscript{2}s\textsuperscript{-1} and $T$ between $15$ and $120$ s. For this parameter study our scale heights are fixed at $H = 20$ Mm and $H_\rho = 100$ Mm as these values provide the strongest wave damping of all the values we have considered.

\begin{figure}
  \includegraphics[width=\linewidth]{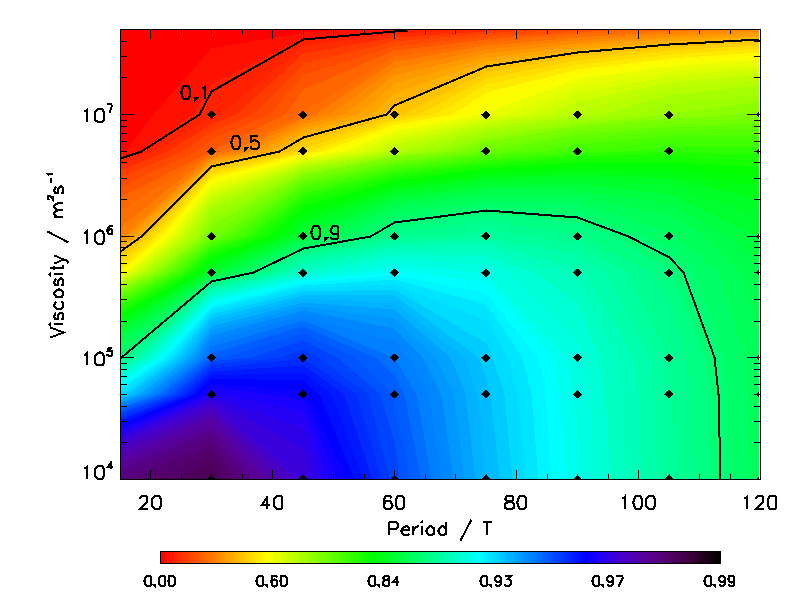}
\caption{Contour plot showing values of the normalised wave energy flux, $\Pi(z_{\text{max}})/\Pi(0)$, from our viscosity-period parameter study. Note that the colour scale is logarithmic. The wave period $T$ is on the $x$-axis and the kinematic viscosity $\nu$ is on the $y$-axis. The diamond symbols represent data points from the parameter study and the thick line contours indicate where $\Pi(z_{\text{max}})/\Pi(0)$ has values of 0.1, 0.5 and 0.9.}
\label{fig:visc_period_grid}
\end{figure}

In \cref{fig:visc_period_grid} we see a contour plot of $\Pi(z_{\text{max}})/\Pi(0)$ for the different values of $\nu$ and $T$ used in this study. This plot is particularly good at demonstrating the different dependencies of phase mixing and wave reflection. Phase mixing is stronger for shorter period waves and higher viscosities whilst wave reflection is stronger for longer period waves and is independent of viscosity.

We can see that for higher viscosities where viscous dissipation through phase mixing is the dominant cause of wave damping, the damping decreases as the period is increased because of the reduced effect of phase mixing. In contrast for low viscosities where there is very little wave damping caused by phase mixing, we can see that wave damping increases as the period is increased due to the growing effect of wave reflection.

We can also see that in the short period limit where phase mixing is strong and wave reflection is weaker there is a very strong dependence of the wave energy on viscosity. In the long period limit, however, the wave damping is mostly independent of the viscosity as it is caused primarily by wave reflection. 

\section{Finite difference solver: Wigglewave}
\label{sec:wigglewave}

\begin{figure*}
  \includegraphics[width=\linewidth]{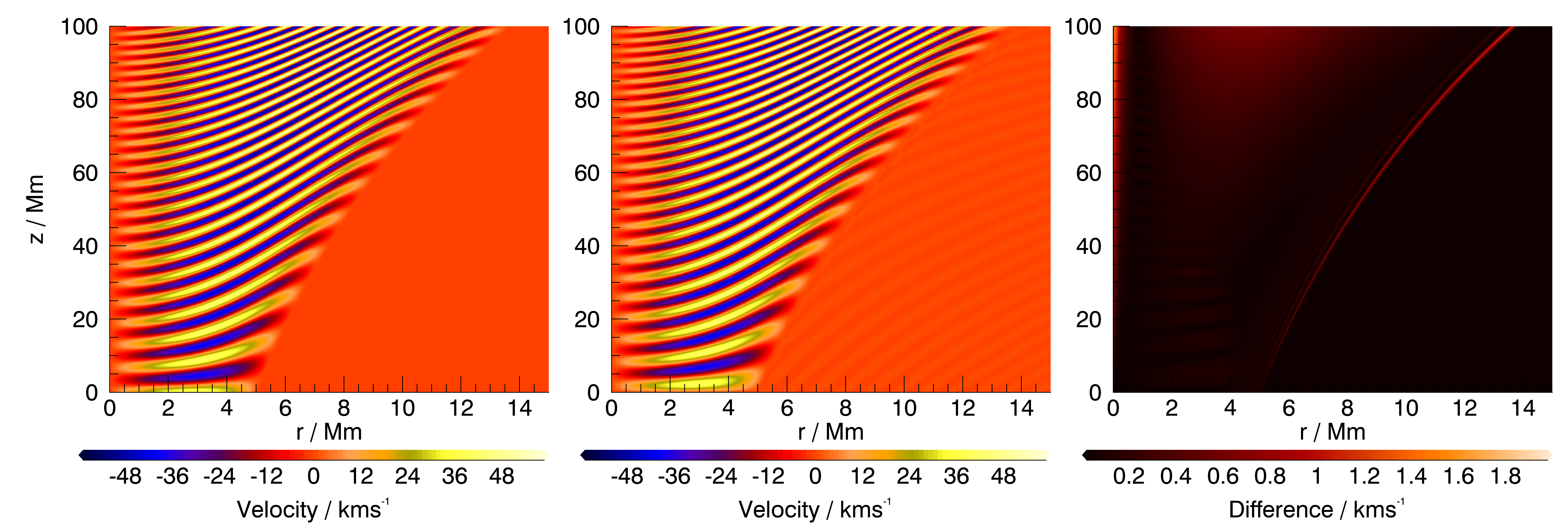}
\caption{Contour plots of the velocity profile for the torsional Alfv\'en wave, in the case where $T=10$ s, as calculated using \textit{TAWAS} (left) and simulated using \textit{WiggleWave} (centre) as well as a contour plot of the difference between the envelopes of these two wave profiles (right).}
\label{fig:velocity_compare}
\end{figure*}

\begin{figure*}
  \includegraphics[width=\linewidth]{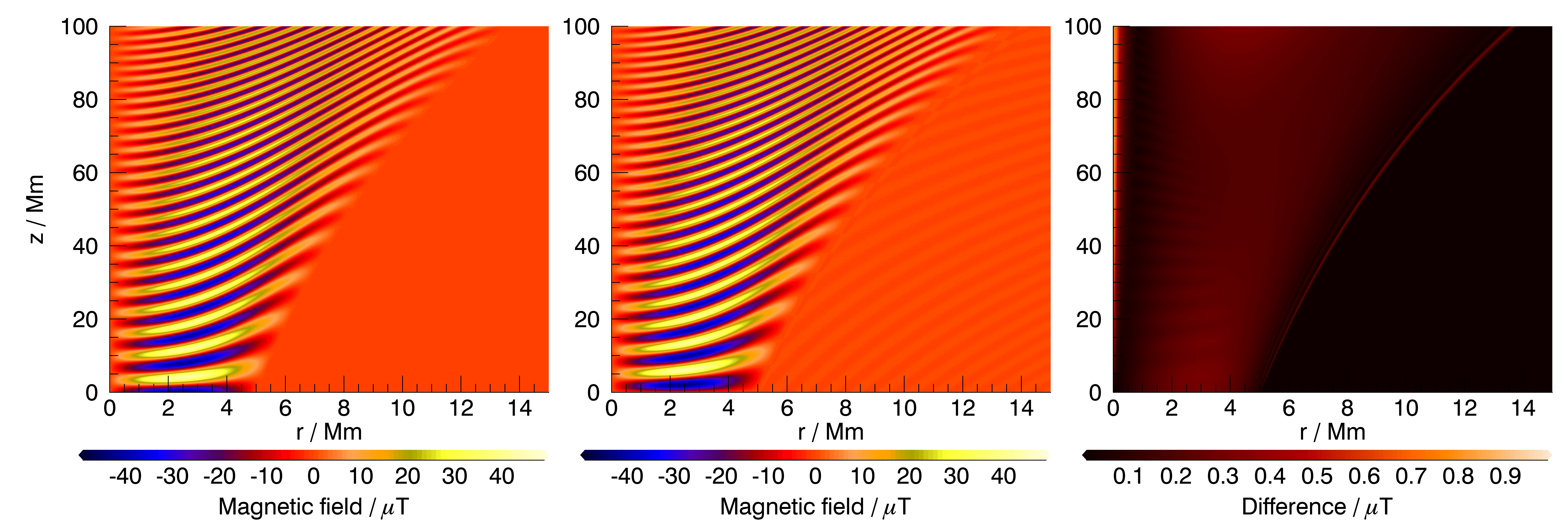}
\caption{Contour plots of the magnetic field profile for the torsional Alfv\'en wave, in the case where $T=10$ s, as calculated using \textit{TAWAS} (left) and simulated using \textit{WiggleWave} (centre) as well as a contour plot of the difference between the envelopes of these two wave profiles (right).}
\label{fig:magnetic_compare}
\end{figure*}

Now that we have determined the parameter space for which wave damping is the strongest we can test the validity of the analytical solution both within and beyond the limits of the WKB approximation used in its derivation. We will do this by comparing the results for the Alfv\'en wave propagation calculated in \textit{TAWAS} to the results from \textit{WiggleWave}, a FORTRAN code written to solve the linearised governing equations for this system, \cref{eq:velocity,eq:magnetic}, directly using finite differencing methods. 

\textit{WiggleWave} uses a fourth-order central difference scheme for calculating spatial derivatives and a fourth-order Runge-Kutta (RK4) scheme for updating at each timestep. \textit{WiggleWave} calculates solutions for $v$ and $b$ over two-dimensional grid in radius $r$ and height $z$, the wave envelopes for $v$ and $b$ are also calculated. The source code and more details for \textit{WiggleWave} can be found at \url{https://github.com/calboo/Wigglewave}.

For simulations in \textit{WiggleWave} the grid resolution used over the domain was the same as in \textit{TAWAS}, that is to $500\times2000$, in the $r$ and $z$ directions respectively. Although exponential damping regions were also included for the upper vertical and outer radial boundaries when necessary, to prevent wave reflection from these boundaries. The numerical setup in \textit{WiggleWave} used was identical to that used in \textit{TAWAS} and described in \cref{sec:setup} except that wave driving at the lower boundary was prescribed within the tube boundary $r = r_0$ as,

\begin{align}
v \quad = \quad &
    u_0\frac{r}{r_0}\left(1-\left(\frac{r}{r_0}\right)^2\right)
    \sin\left(\omega\left(\frac{z}{V_A}-t\right)\right), \\
b \quad = \quad &
    -u_0\frac{r}{r_0}\left(1-\left(\frac{r}{r_0}\right)^2\right)
    \sin\left(\omega\left(\frac{z}{V_A}-t\right)\right) \sqrt{\mu_0\rho}    
\end{align}
where we have approximated the lower boundary as a flat magnetic surface. 

First, we compare the results of \textit{TAWAS} and \textit{WiggleWave} in a scenario where wave damping is relatively weak so that we can compare the results for cases in which the WKB approximation is valid. We set the magnetic scale height to $H = 50$ Mm and a density scale height to $H_\rho = 50$ Mm, the kinematic viscosity is set to $\nu = 5 \times 10^7$ m\textsuperscript{2}s\textsuperscript{-1}. We consider cases with different wave periods, $T \in [10,30,60,90,120]$ s. Each case in this scenario was run for 3000 s to allow the wave to fully propagate throughout the domain and reach a state of quasi-equilibrium.

The data in \cref{fig:velocity_compare,fig:magnetic_compare} is from the case where $T = 10$ s. \cref{fig:velocity_compare} shows contour plots of the velocity perturbations from both \textit{TAWAS} and \textit{WiggleWave} as well as the difference between the envelopes of these two wave profiles. We can see that, ignoring the phase difference, the wave contours are almost identical. The maximum wave amplitude is about 60 km s\textsuperscript{-1} whilst the difference between the wave envelopes is well below 1 km s\textsuperscript{-1} over most of the domain. Similarly \cref{fig:magnetic_compare} shows contour plots of the magnetic field perturbations from both \textit{TAWAS} and \textit{WiggleWave} as well as the difference between the envelopes of these two wave profiles. Again we can see that the wave contours are almost identical. The maximum wave amplitude is about 50 $\mu T$ whilst the difference between the wave envelopes is below 0.5 $\mu T$ over most of the domain.

\begin{figure}
  \includegraphics[width=\linewidth]{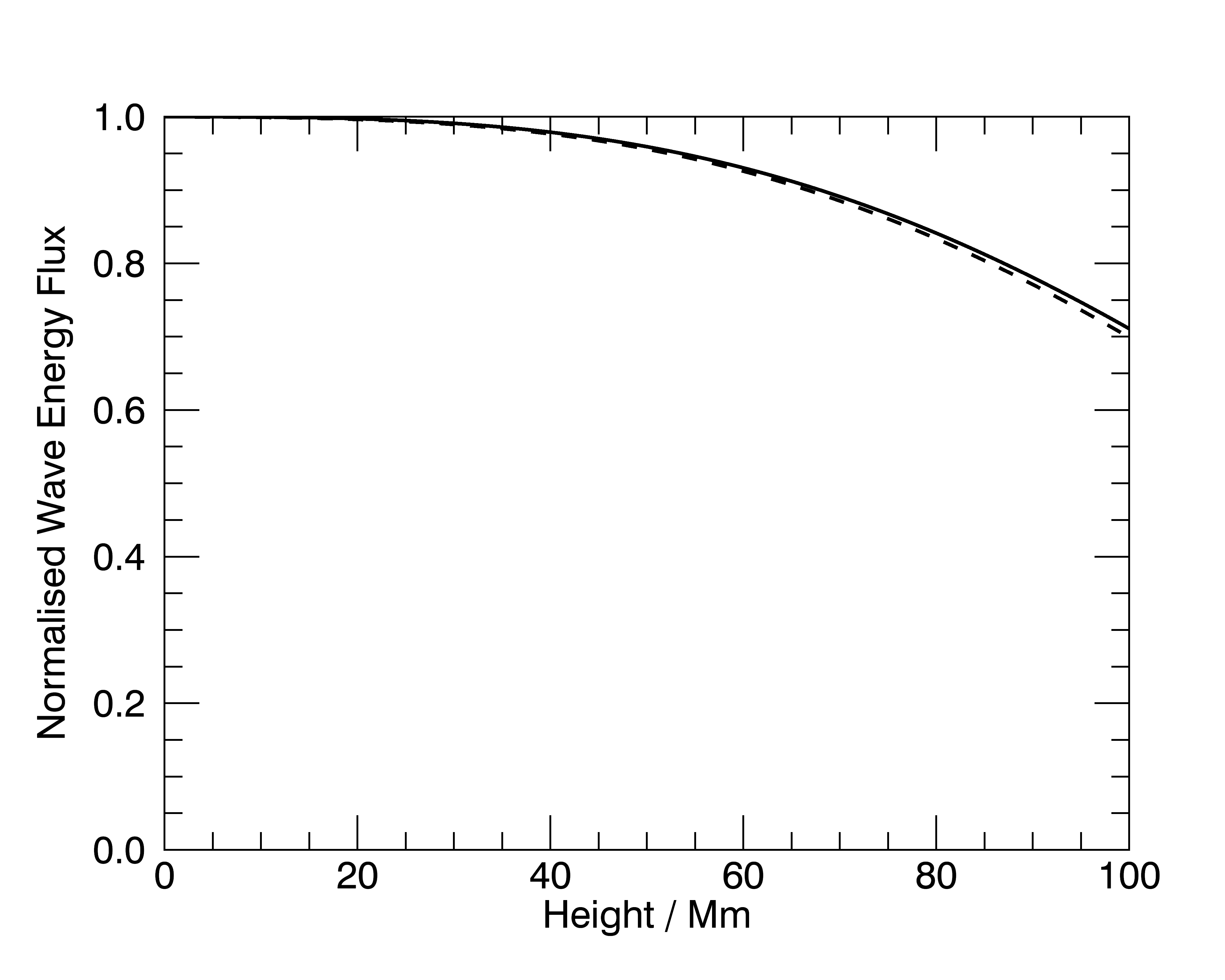}
\caption{Graph of the normalised wave energy flux across magnetic surfaces plotted against the height at which each magnetic surface intersects the $z-$axis. This graph is for a scenario with $H = 50$ Mm, $H_\rho = 50$ Mm, $\nu = 5 \times 10^7$ m\textsuperscript{2}s\textsuperscript{-1} and $T = 10$ s. The wave energy flux from \textit{TAWAS} is shown as a solid  line and the wave energy flux calculated from the \textit{WiggleWave} solution is shown as a dashed line.}
\label{fig:power_B10}
\end{figure}

We can also compare graphs of the wave energy flux $\Pi(z)$ across magnetic surfaces of increasing height. \cref{fig:power_B10} shows graphs of the normalised wave energy flux against the height at which each magnetic surface intersects the $z-$axis for the case where $T = 10$ s. We can see that although the curves from \textit{TAWAS} and \textit{WiggleWave} show significant damping at higher heights due to enhanced phase mixing, the difference between the two curves is negligible. In fact, the maximum difference in wave energy flux for this case is only $1.2\%$. For cases with longer wave periods the damping is much less and the difference between the wave energy flux curves is always less than $1\%$. This shows that within the limits of the WKB approximation the analytical model provides a valid solution to the linear governing equations.

Now we consider a different scenario to test the predictions of the analytical solution beyond the limits of the WKB approximation. We use a magnetic scale height of $H = 20$ Mm and a density scale height of $H_\rho = 100$ Mm to maximise wave damping. Again we set the kinematic viscosity to $\nu = 5 \times 10^7$ m\textsuperscript{2}s\textsuperscript{-1} and consider cases with different wave periods, $T \in [10,30,60,90,120]$ s. Each case in this scenario was run for 4500 s to allow the wave to fully propagate throughout the domain and reach a state of quasi-equilibrium.

When comparing the wave envelopes for this scenario it is immediately obvious that there are differences between the \textit{TAWAS} and \textit{WiggleWave} solutions in all cases. This shows that indeed the WKB approximation used in the analytical solution is no longer valid when wave damping is strong, as we would expect. The question then is whether the wave damping and crucially the viscous damping, which is relevant to coronal heating, is stronger or weaker than the analytical model predicts.

\begin{figure}
  \includegraphics[width=\linewidth]{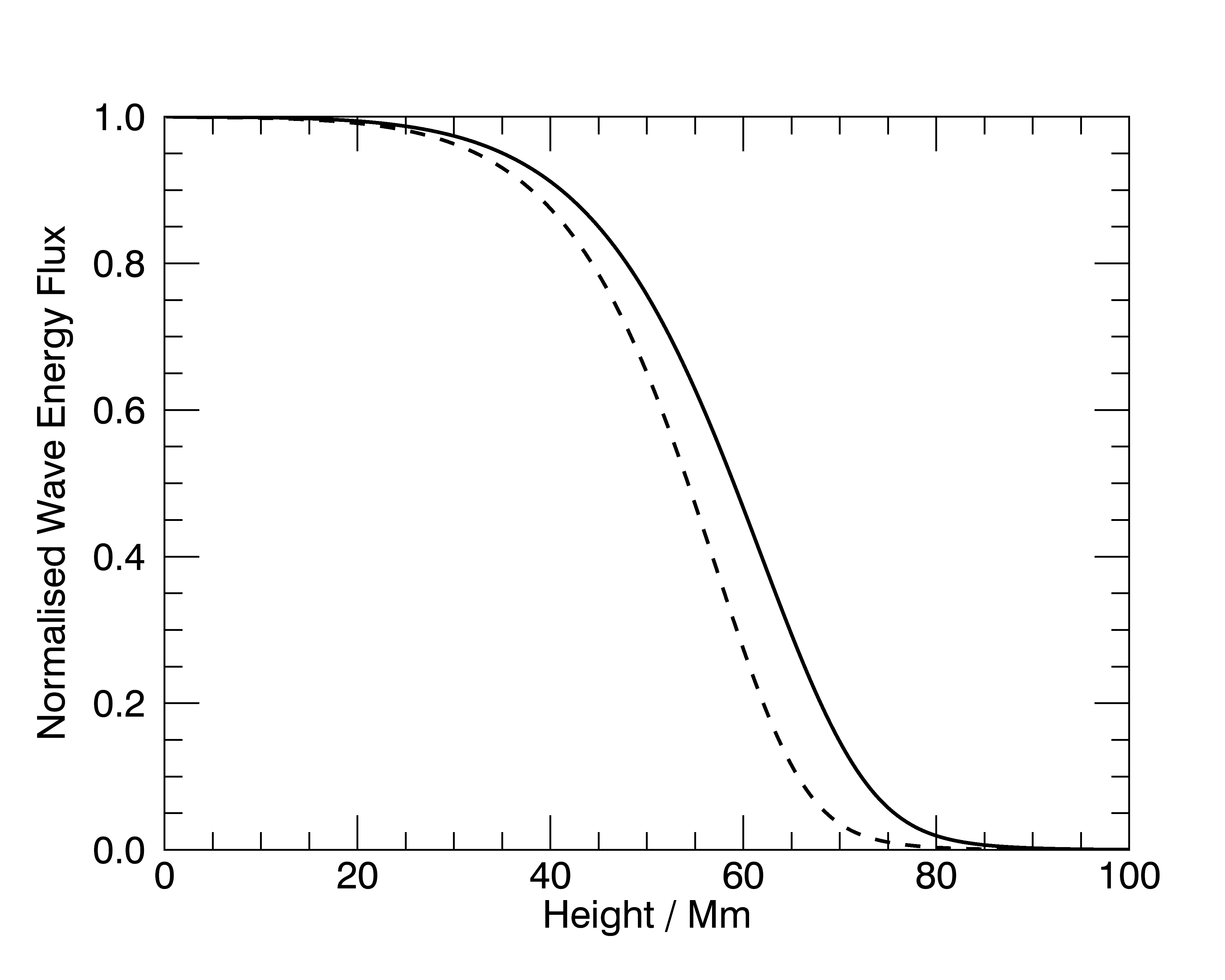}
\caption{Graph of the normalised wave energy flux across magnetic surfaces plotted against the height at which each magnetic surface intersects the $z-$axis. This graph is for a scenario with $H = 20$ Mm, $H_\rho = 100$ Mm, $\nu = 5 \times 10^7$ m\textsuperscript{2}s\textsuperscript{-1} and $T = 10$ s. The wave energy flux from \textit{TAWAS} is shown as a solid  line and the wave energy flux calculated from the \textit{WiggleWave} solution is shown as a dashed line.}
\label{fig:power_C10}
\end{figure}

\begin{figure}
  \includegraphics[width=\linewidth]{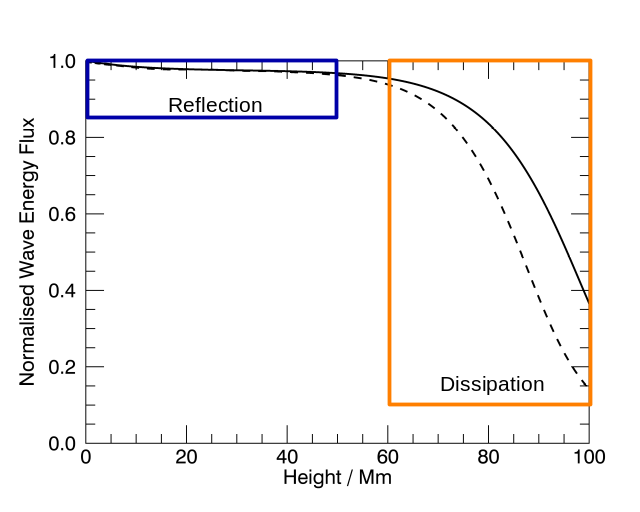}
\caption{Graph of the normalised wave energy flux across magnetic surfaces plotted against the height at which each magnetic surface intersects the $z-$axis. This graph is for a scenario with $H = 20$ Mm, $H_\rho = 100$ Mm, $\nu = 5 \times 10^7$ m\textsuperscript{2}s\textsuperscript{-1} and $T = 60$ s. The wave energy flux from \textit{TAWAS} is shown as a solid  line and the wave energy flux calculated from the \textit{WiggleWave} solution is shown as a dashed line. The annotated boxes shows regions where reflective damping (blue box) or viscous dissipation (orange box) are the dominant damping mechanism.}
\label{fig:power_C60}
\end{figure}

To answer this question we can compare the graphs of the wave energy flux $\Pi(z)$ for these cases, \cref{fig:power_C10} shows a graph of the normalised wave energy flux for the case where $T = 10$ s and \cref{fig:power_C60} for the case where $T = 60$ s. We can see in \cref{fig:power_C10} that in the case of $T = 10$ s the wave is fully damped before it reaches 100 Mm for both solutions, we can also see that the \textit{WiggleWave} solution shows stronger damping that the analytic solution from \textit{TAWAS}. We can see in \cref{fig:power_C60} that in the case of $T = 60$ s the wave is strongly damped at 100 Mm height but much more strongly in the \textit{WiggleWave} solution than in the analytic solution from \textit{TAWAS}. This demonstrates that beyond the WKB limit the analytic solution actually under-predicts the wave damping. As a side note, this graph also shows the effect of wave reflection in the lower domain for both solutions.

We can see in \cref{fig:power_C10} that for the case of $ T= 10$ s the wave damping is caused almost entirely by viscous dissipation. Looking at \cref{fig:power_C60}, however, we can see that for the case of $T = 60$ s the effect of wave reflection of wave energy flux is visible in the lower part of the domain for both the \textit{TAWAS} and \textit{WiggleWave} results. This is because the longer period wave has a larger wavelength and is  therefore more prone to reflection, although as the wave propagates its wavelength decreases and so does the effect of wave reflection. It is interesting to note that solutions only diverge when viscous damping becomes strong due to enhanced phase mixing higher up in the domain.

\section{Conclusions}
\label{sec:conclusions}
 
 In this paper we analytically and numerically studied the enhanced phase mixing of torsional Alfv\'en waves in an expanding magnetic flux tube embedded within a stratified coronal atmosphere. This type of structure is typical of a coronal plume or divergent coronal loop. We began in \cref{sec:analytic_solutions} by formulating an analytical solution to the linearised governing equations \cref{eq:velocity,eq:magnetic} by following the derivation in \citep{Ruderman2018}, that uses the WKB approximation, and making a correction toward the end of the formulation. 
 
 The IDL code \textit{TAWAS} was presented in \cref{sec:tawas}. \textit{TAWAS} calculates solutions using the analytic formula over a grid in radius, $r$ and height, $z$ as well as the total wave energy flux of the Alfv\'en wave over magnetic surfaces at different heights. In \cref{sec:correction} we showed that it is important to include the effect of reflection when considering Alfv\'en wave damping and that our correction to the analytical solution decreases the amount of damping attributable to wave reflection relative to the uncorrected formula.
 
 Then in \cref{sec:param_studies} we used \textit{TAWAS} (\url{https://github.com/calboo/TAWAS}) to perform two parameter studies to identify the parameter space in which Alfv\'en wave damping is significant over 100 Mm. In the first parameter study we showed that damping is only significant when the magnetic scale height is small $H < 30$ Mm (i.e. for strongly divergent magnetic field lines) and the density stratification is $H_\rho$ is larger. In the second parameter study we showed that damping is stronger for lower period waves and higher kinematic viscosities. 
 
 To test the validity of our analytical solution, both within and beyond the limits of the WKB approximation, we wrote a finite difference solver in FORTRAN called \textit{Wigglewave} (\url{https://github.com/calboo/Wigglewave}) that directly solves the linearised governing equations \cref{eq:velocity,eq:magnetic} and is fourth-order accurate in time and space. In \cref{sec:wigglewave} we compared results from \textit{TAWAS} and \textit{Wigglewave}. We showed that for scenarios with weak damping ($H = 50$ Mm, $H_\rho = 50$ Mm, $\nu = 5 \times 10^7$), the analytical solution matches the exact solution very well (to within 1\% over most of the domain). Whereas for scenarios with strong damping ($H = 20$ Mm, $H_\rho = 100$ Mm) the analytical solution under predicts wave damping compared to the exact solution calculated by \textit{Wigglewave}. As a side note, our \textit{wigglewave} results also showed the effect of wave reflection in graphs of the wave energy flux.
 
 The simulation results from \textit{Wigglewave} show that, assuming a reasonable transverse density gradient across the tube structure, the energy of torsional Alfv\'en waves can be fully dissipated within 100 Mm, if the magnetic field lines are highly divergent, $H \sim 20$ Mm, the wave period is short, $T \sim 10$ s and a high value of anomalous kinematic viscosity, $\nu = 5 \times 10^7$ m\textsuperscript{2}s\textsuperscript{-1}, is assumed. These findings are similar to those in \citep{Smith} which considers shear Alfv\'en waves and shows that shear waves of a similar amplitude are sufficient to satisfy the coronal heating requirement \citep{Aschwanden} when they are fully dissipated in sufficiently divergent active regions. Note, that in these simulations only monochromatic Alfv\'en wave driving has been considered, if instead a broad spectrum driver is considered as in \citep{Tsik_wide} and \citep{Arber} then the heating requirement could potentially be fulfilled for lower values of viscosity.
 
  Furthermore it is stated in \citep{Smith} that additional physical effects, such as pressure, three-dimensionality and non-linearity, would lead to further heat deposition from the Alfv\'en waves. Indeed, in \citep{Malara_3d}, it is shown that three dimensional effects can sometimes dissipate wave energy more efficiently than phase mixing, and in \citep{Malara_Compressible}, it is shown that the interaction of Alfv\'en waves with a inhomogeneity can give rise to compressible perturbations. We will explore these effects further in Paper II using full MHD simulations. To summarise our findings in this paper:
 
 \begin{enumerate}
 
 \item  Wave reflection should be accounted for in the analytical solution for the enhanced phase mixing of torsional Alfv\'en waves. The \textit{Wigglewave} results show that the effect of wave reflection is significant and can be accurately predicted by our corrected analytic formula. \\
 
 \item  The analytical solution for the enhanced phase mixing of torsional Alfv\'en waves is accurate at predicting the wave energy flux of the Alfv\'en wave within the limits of the WKB approximation but under-reports wave damping beyond the WKB limit. \\
 
 \item  The energy dissipated due to the enhanced phase mixing of torsional Alfv\'en waves is enough to fulfill the coronal heating requirement in a magnetic field with sufficiently divergent field lines, although this requires the application of anomalous viscosity and the presence of high frequency Alfv\'en waves.
 
 \end{enumerate}

\section*{Acknowledgements}

C.B. would like to thank UK STFC DISCnet for financial support of his PhD studentship. This research utilized Queen Mary's Apocrita HPC facility, supported by QMUL Research-IT \url{http://doi.org/10.5281/zenodo.438045}.

\section*{Data Availability}

All data used in this study was generated by either our analytical numerical solver \textit{TAWAS},
\url{https://github.com/calboo/TAWAS}, or our finite difference solver \textit{Wigglewave}, \url{https://github.com/calboo/Wigglewave}.



\bibliographystyle{mnras}
\bibliography{Enhanced_Phase_Mixing_Paper_I} 

\newpage
 \appendix
 \section{Solution for a Potential Axisymmetric Magnetic Field}
 \label{App1}
 
 We require that the background magnetic field for our setup $\mathbf{B_0}$ is a potential field, ($\nabla^2\mathbf{B_0} = 0$), is axisymmetric, ($\partial_\theta = 0$), has no azimuthal component ($\mathbf{B_0} = (B_r,0,B_z)$) and decays with height according to a characteristic scale height $H$.We begin by considering how the Laplace operator works in cylindrical coordinates for a random vector $\mathbf{A}$ and random function $f$,
 
 \begin{equation}
\nabla^2\mathbf{A}= \begin{pmatrix}
					\displaystyle{\nabla^2 A_r - \frac{A_r}{r^2} - \frac{2}{r^2}\diffp{A_\theta}{\theta}} \\
					\displaystyle{\nabla^2 A_\theta - \frac{A_\theta}{r^2} + \frac{2}{r^2}\diffp{A_r}{\theta}} \\
					\displaystyle{\nabla^2 A_z},
					\end{pmatrix}
\end{equation}
 \begin{equation}
\nabla^2 f=  \left(\frac{1}{r}\diffp{}{r}\left(r\diffp{f}{r}\right) + \frac{1}{r^2}\diffp[2]{f}{\theta} + \diffp[2]{f}{z} \right).
\end{equation}
Now substituting $\mathbf{B_0}$ into our Laplace equation and applying our assumptions of axisymmetry and no azimuthal component we have equations for $B_r$ and $B_z$, 

 \begin{equation}
\nabla^2\mathbf{\mathbf{B_0}}= \begin{pmatrix}
					\displaystyle{\frac{1}{r}\frac{\partial}{\partial r}\left(r \diffp{B_r}{r}\right)
                    + \diffp[2]{B_r}{z}
                    -\frac{B_r}{r^2}} \\
					\displaystyle{0} \\
					\displaystyle{\frac{1}{r}\frac{\partial}{\partial r}\left(r \diffp{B_z}{r}\right)
                    + \diffp[2]{B_z}{z}}
					\end{pmatrix} = \mathbf{0}.
\end{equation}

\subsection{Solution for \texorpdfstring{$B_r$}{Br}}

Let us first consider the solution for $B_r$,

\begin{equation}
\frac{1}{r}\frac{\partial}{\partial r}\left(r \diffp{B_r}{r}\right)
+ \diffp[2]{B_r}{z}
-\frac{B_r}{r^2}
= \;0 .
\end{equation}
We try a separable solution $B_r= Z_r(z)R_r(r)$ which gives us,

\begin{equation}
Z_r\frac{1}{r}\frac{\partial}{\partial r}\left(r \diffp{R_r}{r}\right)
+ R_r\diffp[2]{Z_r}{z}
-\frac{R_r Z_r}{r^2}
= \; 0,
\end{equation}
\begin{equation}
Z_r\left(\diffp[2]{R_r}{r} + \frac{1}{r}\diffp{R_r}{r}\right)
+ R_r\diffp[2]{Z_r}{z}
-\frac{R_r Z_r}{r^2}
= \;0,
\end{equation}
\begin{equation}
\frac{1}{R_r}\left(\diffp[2]{R_r}{r} + \frac{1}{r}\diffp{R_r}{r}\right)
+ \frac{1}{Z_r}\diffp[2]{Z_r}{z}
-\frac{1}{r^2}
= \;0,
\end{equation}
\begin{equation}
\frac{1}{R_r}\left(\diffp[2]{R_r}{r} + \frac{1}{r}\diffp{R_r}{r}\right)
-\frac{1}{r^2}
= \; -\frac{1}{Z_r}\diffp[2]{Z_r}{z}.
\end{equation}
As both sides are in separate variables they must each be equal to constant for this equality to hold. We call this constant $-k_1^2$, then we have, 

\begin{equation}
\frac{1}{R_r}\left(\diffp[2]{R_r}{r} + \frac{1}{r}\diffp{R_r}{r}\right)
-\frac{1}{r^2} = \; -k_1^2,
\end{equation}
\begin{equation}
-\frac{1}{Z_r}\diffp[2]{Z_r}{z} = \; -k_1^2.
\end{equation}
Note that we have used a $-k_1^2$ instead of $k_1^2$ so that we will end up with exponential solutions for $Z_r$ instead of oscillating solutions. The equation for $Z_r$ is then,

\begin{equation}
\diffp[2]{Z_r}{z} = \; k_1^2 Z_r,
\end{equation}
The only solution for $Z_r$ that decays with characteristic scale height $H$ is,

\begin{equation}
Z_r = e^{-z/H}.
\end{equation}
This allows us to calculate $k_1$,

\begin{equation}
\diffp[2]{Z_r}{z} = \; \frac{1}{H^2} Z_r \quad \implies \quad k_1 = \; \frac{1}{H}.
\end{equation}
The equation for $R_r$ is,

\begin{equation}
\diffp[2]{R_r}{r} + \frac{1}{r}\diffp{R_r}{r}
-\frac{1}{r^2} = \; -k_1^2 R_r,
\end{equation}
\begin{equation}
\diffp[2]{R_r}{r} + \frac{1}{r}\diffp{R_r}{r}
+ \left(k_1^2 - \frac{1}{r^2}\right)R_r = \; 0, 
\end{equation}
now we define $p_1 = k_1r$ and transform our equation so that it is in terms of $p_1$,

\begin{equation}
\diffp[2]{R_r}{p_1} + \frac{1}{r}\diffp{R_r}{p_1}
+ \left(1 - \frac{1}{p_1^2}\right)R_r = \; 0,
\end{equation}
we can write this as,

\begin{equation}
\diffp[2]{R_r}{p_1} + \frac{1}{r}\diffp{R_r}{p_1}
+ \left(1 - \frac{m^2}{p_1^2}\right)R_r = \; 0,
\end{equation}
where m = 1. The solution to this is a Bessel function of the first kind with $m = 1$,

\begin{equation}
R_r = J_1(p_1) = J_1(k_1r) = J_1\left(\frac{r}{H}\right).
\end{equation}
The solution for $B_r$ is therefore,

\begin{equation}
B_r = B_0 e^{-z/H} J_1\left(\frac{r}{H}\right).
\end{equation}

\subsection{Solution for \texorpdfstring{$B_z$}{Bz}}

Now we will consider the solution for $B_z$,

\begin{equation}
\frac{1}{r}\frac{\partial}{\partial r}\left(r \diffp{B_z}{r}\right)
+ \diffp[2]{B_z}{z}
= \; 0.
\end{equation}
We try a separable solution $B_z= Z_z(z)R_z(r)$ which gives us,

\begin{equation}
Z_z\frac{1}{r}\frac{\partial}{\partial r}\left(r \diffp{R_z}{r}\right)
+ R_z\diffp[2]{Z_z}{z}
= \;0,
\end{equation}
\begin{equation}
Z_z\left(\diffp[2]{R_z}{r} + \frac{1}{r}\diffp{R_z}{r}\right)
+ R_z\diffp[2]{Z_z}{z}
= \;0,
\end{equation}
\begin{equation}
\frac{1}{R_z}\left(\diffp[2]{R_z}{r} + \frac{1}{r}\diffp{R_z}{r}\right)
+ \frac{1}{Z_z}\diffp[2]{Z_z}{z}
= \;0, 
\end{equation}
\begin{equation}
\frac{1}{R_z}\left(\diffp[2]{R_z}{r} + \frac{1}{r}\diffp{R_z}{r}\right)
= \; -\frac{1}{Z_z}\diffp[2]{Z_z}{z}.
\end{equation}
As both sides are in separate variables they must each be equal to constant for this equality to hold. We call this constant $-k_2^2$, then we have, 

\begin{equation}
\frac{1}{R_z}\left(\diffp[2]{R_z}{r} + \frac{1}{r}\diffp{R_z}{r}\right)
= \; -k_2^2, 
\end{equation}
\begin{equation}
-\frac{1}{Z_z}\diffp[2]{Z_z}{z} = \; -k_2^2.
\end{equation}
Note that we have used a $-k_2^2$ instead of $k_2^2$ so that we will end up with exponential solutions for $Z_z$ instead of oscillating solutions. The equation for $Z_z$ is then,

\begin{equation}
\diffp[2]{Z_z}{z} = \; k_2^2 Z_z.
\end{equation}
The only solution for $Z_z$ that decays with characteristic scale height $H$ is,

\begin{equation}
Z_z = e^{-z/H}.
\end{equation}
This allows us to calculate $k_2$ which turns out to be the same as $k_1$,

\begin{equation}
\diffp[2]{Z_z}{z} = \; \frac{1}{H^2} Z_z \quad \implies \quad k_2 = \; \frac{1}{H}.
\end{equation}
The equation for $R_z$ is,

\begin{equation}
\diffp[2]{R_z}{r} + \frac{1}{r}\diffp{R_z}{r}
 = \; -k_2^2 R_z,
\end{equation}
\begin{equation}
\diffp[2]{R_z}{r} + \frac{1}{r}\diffp{R_z}{r}
+ k_2^2 R_z = \; 0,
\end{equation}
now we define $p_2 = k_2r$ and transform our equation so that it is in terms of $p_2$,

\begin{equation}
\diffp[2]{R_z}{p_2} + \frac{1}{r}\diffp{R_z}{p_2}
+ R_z = \; 0,
\end{equation}
we can write this as,

\begin{equation}
\diffp[2]{R_z}{p_2} + \frac{1}{r}\diffp{R_z}{p_2}
+ \left(1 - \frac{m^2}{p_2^2}\right)R_z = \; 0,
\end{equation}
where m = 0. The solution to this is a Bessel function of the first kind with $m = 0$,

\begin{equation}
R_z = J_0(p_2) = J_0(k_2r) = J_0\left(\frac{r}{H}\right).
\end{equation}
The solution for $B_z$ is therefore,

\begin{equation}
B_z = B_0 e^{-z/H} J_0\left(\frac{r}{H}\right).
\end{equation}

\subsection{Solution for \texorpdfstring{$\mathbf{B_0}$}{B0}}

The solution for the potential, axisymmetric magnetic field that decays with scale height $H$ is therefore,

\begin{equation}
\mathbf{B_0} = \begin{pmatrix}
B_0 e^{-z/H} J_1\left(\frac{r}{H}\right)\\
0 \\
B_0 e^{-z/H} J_0\left(\frac{r}{H}\right)
\end{pmatrix},
\end{equation}
where $B_0$ is the characteristic magnetic field strength.

\bsp	
\label{lastpage}
\end{document}